\documentclass[showpacs,preprintnumbers,amsmath,amssymb]{revtex4}

\usepackage{epsfig}
\usepackage{graphicx}
\usepackage{dcolumn}
\usepackage{bm}

\begin{document}


\title{Finite-Energy Pseudoparticle Theory for the 1D Hubbard Model
II: Holon and Spinon Dominant Processes for the Few-Electron
Spectral Weight Properties}
\author{J. M. P. Carmelo}
\affiliation{GCEP-Center of Physics, University of Minho, Campus
Gualtar, P-4710-057 Braga, Portugal}
\author{K. Penc}
\affiliation{Research Institute for Solid State Physics and
Optics, H-1525 Budapest, P.O.B. 49, Hungary}
\date{19 November 2002}

\begin{abstract}
In the first paper of this series it was found that the
$\eta$-spin $1/2$ holons, spin $1/2$ spinons, and $c$
pseudoparticles whose occupancy configurations describe the energy
eigenstates of the one-dimensional Hubbard model emerge from the
electron - rotated-electron unitary transformation. An important
breakthrough is that the theory provides relevant information
about the relationship of the original electrons to these quantum
objects. In this second paper we discuss and clarify how such a
relation can be used in a program for evaluation of finite-energy
few-electron spectral functions. As a first step, here we
characterize the dominant holon and spinon microscopic physical
processes that originate more than 99\% of the few-electron
spectral weight. These dominant processes are related to exact
selection rules for the values of the number of holons and spinons
generated or annihilated by application onto a ground state of
rotated-electron operators. While our theory also describes the
higher-order microscopic processes associated with the remaining
less than 1\% few-electron spectral weight, the clarification and
study of the above dominant processes is valuable and useful for
the further understanding and description of the few-electron
spectral properties observed in real low-dimensional materials.
Moreover, in this paper generalize the concepts of a lower Hubbard
band and upper Hubbard bands to all values of on-site Coulombian
repulsion.
\end{abstract}


\maketitle
\section{INTRODUCTION}

In this paper and its two companion papers \cite{I,III} the
relation of the electrons of the one-dimensional (1D) Hubbard
model to the quantum objects whose occupancy configurations
describe its energy eigenstates is investigated for the whole
Hilbert space. The study of such a non-perturbative relation is a
necessary step for the description of the finite-energy
few-electron spectral properties of the many-electron quantum
problem.

The study of the one-dimensional (1D) Hubbard model
\cite{Lieb,Takahashi,HL,Yang89,Martins98,II,Rasetti} for
interacting electrons is of general importance because the
understanding of correlated systems and of their finite-energy
excitations is still far from complete. The problem of the 1D
Hubbard model in the limit of large and infinite on-site Coulomb
repulsion was previously studied in the literature by many authors
\cite{Harris,Mac,Eskes,Geb,Beni,Klein,Carmelo88,Ogata,Parola,Ricardo,Penc95,Penc96,Penc97}.
Our studies of the model for finite values of the on-site
repulsion $U$ are motivated by the anomalous one-electron and
two-electron spectral properties observed in metallic and
insulating phases of quasi-one-dimensional materials, which cannot
be described by the usual Fermi-liquid theory \cite{Pines,Baym}.
Recently there has been a renewed experimental interest on the
properties of these materials
\cite{Hussey,Menzel,Fuji02,Hasan,Ralph,Gweon,Monceau,Takenaka,Mizokawa,Moser,Mihaly,Vescoli00,Denlinger,Fuji,Kobayashi,Bourbonnais,Vescoli,Zwick,Neudert,Mori,Kim}.
Some of these experimental studies observed unusual
finite-energy/frequency spectral properties
\cite{Hasan,Ralph,Gweon,Takenaka,Vescoli}. Since in the case of
finite-excitation energy the Luttinger liquid description does not
apply \cite{Luther,Solyom,Haldane}, these finite-energy/frequency
spectral properties are far from being well understood. However,
there are indications that electronic correlation effects might
play an important role in the finite-energy physics of these
low-dimensional materials
\cite{Hasan,Ralph,Gweon,Takenaka,Vescoli}. For low values of the
excitation energy the microscopic electronic properties of these
materials are usually described by systems of coupled chains. On
the other hand, for finite values of the excitation energy larger
than the transfer integrals for electron hopping between the
chains, 1D lattice models taking into account the screened
electron-electron Coulomb repulsion are expected to provide a good
description of the physics of these materials. The simplest of
these models is the above 1D Hubbard model
\cite{Lieb,Takahashi,Martins98,Rasetti}, which describes such
electron-electron interactions by an effective on-site Coulomb
repulsion $U$. This model corresponds to a non-perturbative
electronic problem. According to the results of the companion
paper \cite{I}, its energy eigenstates can be described by
occupancy configurations of holons, spinons, and $c$
pseudoparticles. Holons and spinons has also been studied for
other models \cite{Faddeed,PWA,Talstra,Tohyama}. The 1D Hubbard
model is often considered a suitable model for the description of
the electronic correlation effects and the non-perturbative
microscopic mechanisms behind the unusual few-electron spectral
properties observed in quasi-one-dimensional materials
\cite{Hasan,Ralph,Vescoli,Mori}. Moreover, recent angle-resolved
ultraviolet photoemission spectroscopy revealed very similar
spectral fingerprints from both high-$T_c$ superconductors and
quasi-one-dimensional compounds \cite{Menzel}. The similarity of
the ultraviolet data for these two different systems could be
evidence of the occurrence of a charge-spin separation associated
with holons and spinons. The anomalous temperature dependence of
the spectral function could also indicate a dimensional crossover
\cite{Menzel,Granath,Orgard,Carlson}. The results of Refs.
\cite{Zaanen,Antonio} also suggest that the unconventional
spectral properties observed in two-dimensional (2D) materials
could have a 1D origin. Thus the holons and spinons could play an
important role in spectral properties of both 1D and 2D
low-dimensional materials.

In this second paper we continue the studies of the first paper of
this series, Ref. \cite{I}. As a preliminary application of the
connection of the concept of rotated electron to the quantum
numbers that label the energy eigenstates provided by the
Bethe-ansatz solution and $\eta$-spin and spin symmetries, in this
paper we use exact holon and spinon selection rules for
rotated-electron operators in the study of the holon and spinon
contents of few-electron excitations. This reveals the dominant
holon and spinon microscopic physical processes that generate more
than 99\% of the spectral weight of few-electron excitations.
While our theory also describes the higher-order processes
associated with the remaining less than 1\% electronic spectral
weight, the clarification of the dominant holon and spinon
microscopic mechanisms is valuable for the further understanding
and description of the few-electron spectral properties observed
in real low-dimensional materials. Fortunately, a preliminary
application of the theoretical predictions which follow from our
study of the holon and spinon contents of few-electron excitations
leads to quantitative agreement with the charge and spin spectral
branch lines observed by means of angle-resolved photoelectron
spectroscopy (ARPES) in the organic conductor TTF-TCNQ
\cite{spectral0}. The preliminary results reported in Ref.
\cite{spectral0} confirm that from the experimental point of view
only the spectral weight associated with the dominant holon and
spinon microscopic processes is observed. Moreover, in this paper
we introduce the concept of an {\it effective electronic lattice}.
The expression of the electrons in terms of holons, spinons, and
pseudoparticles through the electron - rotated-electron
transformation studied here for all values of $U$ is a first
necessary step for the evaluation of finite-energy few-electron
spectral function expressions, as further discussed in Sec. V.

In this paper we also discuss and clarify how the relationship of
the original electrons to the quantum objects whose occupancy
configurations describe the energy eigenstates can be used in a
program for evaluation of finite-energy few-electron spectral
functions. The successful fulfilment of such a program needs the
concepts of local pseudoparticle and effective pseudoparticle
lattice introduced in the third paper of this series, Ref.
\cite{III}. Moreover, in this paper we generalize the concepts of
a lower Hubbard band and upper Hubbard bands to all values of
on-site Coulombian repulsion.

The paper is organized as follows: In Sec. II we provide a short
introduction to the 1D Hubbard model. The definition of the upper
Hubbard bands in terms of rotated-electron double occupation, the
holon and spinon selection rules for rotated-electron operators,
and the concept of an effective electronic lattice are presented
and introduced in Sec. III. The dominant holon and spinon
microscopic physical processes that control the few-electron
spectral properties are studied in Sec. IV. Finally, in Sec. V we
present the discussion of our results and the concluding remarks.
This includes the discussion of the application of the concepts
introduced in the present series of three papers to the fulfilment
of a program for evaluation of finite-energy few-electron spectral
functions.

\section{THE 1D HUBBARD MODEL}

In a chemical potential $\mu $ and magnetic field $H$ the 1D
Hubbard Hamiltonian can be written as,

\begin{equation}
\hat{H} = {\hat{H}}_H  - (U/2)\,[\hat{N} + N_a/2] +
\sum_{\alpha=c,\,s}\mu_{\alpha }\,{\hat{S}}^{\alpha}_z \label{H}
\end{equation}
where,

\begin{equation}
{\hat{H}}_H = \hat{T} + U\,\hat{D} \, , \label{HH}
\end{equation}
is the "simple" Hubbard model. The operators

\begin{equation}
\hat{T} = -t\sum_{j =1}^{N_a}\sum_{\sigma =\uparrow
,\downarrow}\sum_{\delta =-1,+1}
c_{j,\,\sigma}^{\dag}\,c_{j+\delta,\,\sigma} \, , \label{Top}
\end{equation}
and

\begin{equation}
\hat{D} = \sum_{j =1}^{N_a}
\hat{n}_{j,\,\uparrow}\,\hat{n}_{j,\,\downarrow} \, , \label{Dop}
\end{equation}
on the right-hand side of Eq. (\ref{HH}) are the kinetic energy
and the electron double occupation operators, respectively. The
operator

\begin{equation}
\hat{n}_{j,\,\sigma}= c_{j,\,\sigma }^{\dagger }\,c_{j,\,\sigma }
\, , \label{njsig}
\end{equation}
of Eq. (\ref{Dop}) counts the number of spin $\sigma$ electrons at
real-space lattice site $j=1,...,N_a$. The number of lattice sites
$N_a$ is even and large and $N_a/2$ is odd. We consider periodic
boundary conditions. The operators $c_{j,\,\sigma }^{\dagger }$
and $c_{j,\,\sigma}$ which appear in the above equations are the
spin $\sigma $ electron creation and annihilation operators at
site $j$. Moreover, on the right-hand side of Eq. (\ref{H}),
$\mu_c=2\mu$, $\mu_s=2\mu_0 H$, $\mu_0$ is the Bohr magneton, and
the operators ${\hat{S }}^c_z$ and ${\hat{S }}^s_z$ are given in
Eq. (2) of Ref. \cite{I} and are the diagonal generators of the
$\eta$-spin $S^c$ and spin $S^s$ $SU(2)$ algebras
\cite{HL,Yang89,Essler}, respectively. The electron number
operator reads ${\hat{N}}=\sum_{\sigma} \hat{N}_{\sigma}$, where
the operator,

\begin{equation}
{\hat{N}}_{\sigma}=\sum_{j} \hat{n}_{j,\,\sigma} \label{Nsi}
\end{equation}
counts the number of spin $\sigma$ electrons. The momentum
operator reads

\begin{equation}
\hat{P} = - {i\over 2} \sum_{\sigma}\sum_{j=1}^{N_a}\Bigl[c^{\dag
}_{j,\,\sigma}\,c_{j+1,\,\sigma} - c^{\dag
}_{j+1,\,\sigma}\,c_{j,\,\sigma}\Bigr] \, , \label{Popel}
\end{equation}
and commutes with the Hamiltonian introduced in Eq. (\ref{H}).

There are $N_{\uparrow}$ spin-up electrons and $N_{\downarrow}$
spin-down electrons in the chain of $N_a$ sites, lattice constant
$a$, and length $L=[N_a\,a]$ associated with the model (\ref{H}).
We introduce Fermi momenta which in the present thermodynamic
limit $L\rightarrow\infty$ are given by $\pm k_{F\sigma}=\pm \pi
n_{\sigma }$ and $\pm k_F=\pm [k_{F\uparrow}+
k_{F\downarrow}]/2=\pm \pi n/2$, where $n_{\sigma}=N_{\sigma}/L$
and $n=N/L$. The electronic density can be written as
$n=n_{\uparrow }+n_{\downarrow}$ and the spin density is given by
$m=n_{\uparrow}-n_{\downarrow}$. In general we consider densities
in the domains $0\leq n \leq 1/a$\, ; $1/a\leq n \leq 2/a$ and
$-n\leq m \leq n$\, ; $-(2/a-n)\leq m \leq (2/a-n)$, respectively.
However, in the case of the study of transitions whose initial
state is a ground state, for simplicity we restrict our
considerations to values of the electronic density $n$ and spin
density $m$ such that $0\leq n \leq 1/a$ and $0\leq m \leq n$,
respectively. The Hamiltonian ${\hat{H}}_{SO(4)}\equiv {\hat{H}}_H
- (U/2)\,[\hat{N} + N_a/2]$ commutes with the six generators of
the $\eta$ spin $S_c$ and spin $S_s$ algebras
\cite{HL,Yang89,Essler}, their off-diagonal generators being given
in Eqs. (9) and (10) of Ref. \cite{I}. The Bethe-ansatz
solvability of the 1D Hubbard model is restricted to the Hilbert
subspace spanned by the lowest-weight states (LWSs) of the
$\eta$-spin and spin algebras, {\it i.e.} such that $S^{\alpha}=
-S^{\alpha}_z$ \cite{Essler}.

As in the first paper of this series \cite{I}, here we denote the
holons (and spinons) according to their $\eta$-spin projections
$\pm 1/2$ (spin projections $\pm 1/2$). For the definition of the
holon and spinon numbers, their relations, and other aspects of
the holon, spinon, and pseudoparticle description which are useful
for the studies of this paper see the companion paper \cite{I}. An
important concept introduced in Ref. \cite{II} is that of CPHS
ensemble subspace where CPHS stands for $c$ pseudoparticle, holon,
and spinon. This is a Hilbert subspace spanned by all states with
fixed values for the $-1/2$ Yang holon number $L_{c,\,-1/2}$,
$-1/2$ HL spinon number $L_{s,\,-1/2}$, $c$ pseudoparticle number
$N_c$, and for the sets of $\alpha ,\nu$ pseudoparticle numbers
$\{N_{c,\,\nu}\}$ and $\{N_{s,\,\nu}\}$ corresponding to the
$\nu=1,2,3,...$ branches. We note that according to the notation
of Ref. \cite{I} in {\it HL spinon} HL stands for {\it Heilmann
and Lieb}.

\section{THE UPPER HUBBARD BANDS, SELECTION RULES FOR ROTATED-ELECTRON OPERATORS,
AND THE EFFECTIVE ELECTRONIC LATTICE}

In this section we generalize the large-$U$ concept of upper
Hubbard band \cite{Eskes} to all values of the on-site electronic
repulsion $U$ in terms of rotated-electron double occupation.
Moreover, we use the relation of rotated electrons to holons and
spinons to introduce rotated-electron exact selection rules. These
rules limit the values of the deviations in holon and spinon
numbers associated with excited states generated by application of
rotated-electron operators onto any eigenstate of the electron
spin $\sigma$ number operator (\ref{Nsi}). Deviation values
outside the ranges given by these selection rules correspond to
forbidden excited states. In addition, in this section we
introduce the concept of an {\it effective electronic lattice}.
This concept is further used in the third paper of this series,
Ref. \cite{III}, in the introduction of the concepts of a {\it
local pseudoparticle} and an {\it effective pseudoparticle
lattice}. The latter concepts play an important role in the
studies of finite-energy few-electron spectral functions of Refs.
\cite{IIIb,V}.

\subsection{ROTATED ELECTRONS AND THE UPPER HUBBARD BANDS}

As discussed in the companion paper \cite{I}, the operator
${\tilde{c}}_{j,\,\sigma}^{\dag}$ represents the rotated
electrons. The relation of such an operator to the electronic
operator $c_{j,\,\sigma}^{\dag}$ is such that,

\begin{equation}
c_{j,\,\sigma}^{\dag} =
{\hat{V}}(U/t)\,{\tilde{c}}_{j,\,\sigma}^{\dag}\,{\hat{V}}^{\dag}(U/t)
\, . \label{c+tilINV}
\end{equation}
Here ${\hat{V}}(U/t)$ is the electron - rotated-electron unitary
operator uniquely defined by Eqs. (43)-(45) of Ref. \cite{I}. The
rotated-electron double-occupation operator plays an important
role in our studies and is given by,

\begin{equation}
\tilde{D} \equiv {\hat{V}}^{\dag}(U/t)\,\hat{D}\,{\hat{V}}(U/t) =
\sum_{j}\, {\tilde{c}}_{j,\,\uparrow }^{\dagger
}\,{\tilde{c}}_{j,\,\uparrow }\, {\tilde{c}}_{j,\,\downarrow
}^{\dagger }\,{\tilde{c}}_{j,\,\downarrow } \, , \label{Doptil}
\end{equation}
where $\hat{D}$ is the electron double occupation operator given
in Eq. (\ref{Dop}). Note that $c_{j,\,\sigma}^{\dag}$ and
${\tilde{c}}_{j,\,\sigma}^{\dag}$ are only identical in the
$U/t\rightarrow\infty$ limit where electron double occupation
becomes a good quantum number.

Let the operator ${\hat{O}}_{{\cal{N}}}(k)$ (or
${\tilde{O}}_{{\cal{N}}}(k)$) have momentum $k$ and be a product
of a finite number

\begin{equation}
{\cal{N}}=\sum_{l_c,\,l_s=\pm 1} {\cal{N}}_{l_c,\,l_s} \, ,
\label{d}
\end{equation}
of one-electron (or one-rotated-electron) elementary creation
and/or annihilation operators. Here the ratio ${\cal{N}}/N_a$
vanishes in the present thermodynamic limit and
${\cal{N}}_{l_c,\,l_s}$ is the number of electron (or
rotated-electron) creation and annihilation operators for $l_c=-1$
and $l_c=+1$, respectively, and with spin down and spin up for
$l_s=-1$ and $l_s=+1$, respectively.

The studies we present below in this paper refer to the
transformations generated by the application of the general
${\cal{N}}$-electron operator ${\hat{O}}_{{\cal{N}}}(k)$ (or
${\cal{N}}$-rotated-electron operator
${\tilde{O}}_{{\cal{N}}}(k)$) onto any eigenstate of the spin
$\sigma$ electron number operator (\ref{Nsi}). However, we focus
our attention on the particular case of states
${\hat{O}}_{{\cal{N}}}(k)\vert GS\rangle$ (or
${\tilde{O}}_{{\cal{N}}}(k)\vert GS\rangle$) which result from
application of such an operator onto a ground state $\vert
GS\rangle$. We note that the number of spin $\sigma$ electrons
equals the number of spin $\sigma$ rotated electrons.

According to the results of the companion paper \cite{I},
rotated-electron double occupation equals the number of $-1/2$
holons. We emphasize that for electronic densities $n$ such that
$0<n<1/a$ the ground state has no $-1/2$ holons \cite{I,II}. Since
rotated-electron double occupation is a good quantum number, in
this case we can write a general ${\cal{N}}$-electron spectral
function as follows,

\begin{equation}
A_{\cal{N}}(k,\,\omega) =
\sum_{M=0}^{\infty}\,A_{{\cal{N}},M}(k,\,\omega) \, , \label{ONsf}
\end{equation}
where

\begin{equation}
A_{{\cal{N}},M}(k,\,\omega) =\sum_{j}\, \vert\langle M,j\vert
{\hat{O}}_{{\cal{N}}}(k) \vert GS\rangle\vert^2\,\delta\Bigl(
\omega -\omega_{M,j}\Bigr) \, , \label{ONMsf}
\end{equation}
the excitation energy $\omega$ is positive, the $M$ summation of
Eq. (\ref{ONsf}) runs over the values of rotated-electron double
occupation $M=0,1,2,...$ of the excited energy eigenstates $\vert
M,j\rangle$, the $j$ summation of Eq. (\ref{ONMsf}) runs over all
available excited energy eigenstates $\vert M,j\rangle$ with the
same value $M$ of rotated-electron double occupation, and
$\omega_{M,j}\equiv [E_{M,j}-E_{GS}]$ are the excitation energies
relative to the initial ground state. Examples of few-electron
operators ${\hat{O}}_{{\cal{N}}}(k)$ whose spectral-function
excitation energy is usually considered positive are the
one-electron spin $\sigma$ creation operator of momentum $k$, the
two-electron charge operator of momentum $k$, and the two-electron
singlet or triplet Cooper pair addition operator of momentum $k$.

We call {\it lower Hubbard band} the spectral weight distribution
associated with the function $A_{{\cal{N}},0}(k,\,\omega)$ defined
in Eq. (\ref{ONMsf}). Moreover, we call {\it $Mth$ upper Hubbard
band} the spectral weight distribution associated with the
function $A_{{\cal{N}},M}(k,\,\omega)$ defined in Eq.
(\ref{ONMsf}) such that $M>0$. In the ensuing section we find that
for one-electron and two-electron operators most spectral weight
correspond to the lower Hubbard band and first (or first and
second) upper Hubbard band(s). The detailed study of the
few-electron spectral functions (\ref{ONsf}) is for finite values
of $U$ a complex many-electron open problem. Such a problem is
studied in Ref. \cite{V} by the combination of the pseudofermion
representation introduced in Ref. \cite{IIIb} with the holon and
spinon description introduced and studied in this paper and in its
companion papers \cite{I,III}, as discussed in Sec. V. In this
paper we use the relation of electrons and rotated electrons to
the quantum objects whose occupancy configurations describe the
ground state and excited states $\vert M,j\rangle$ that appear in
Eqs. (\ref{ONsf}) and (\ref{ONMsf}) to obtain useful information
about the finite-energy few-electron spectral distributions.

An interesting property is that in the metallic phase of the 1D
Hubbard model the minimum value of the excitation energy $\omega$
for the lower Hubbard band is zero. This justifies the designation
{\it lower band}. On the other hand, the minimum value of the
excitation energy for the $Mth$ upper Hubbard band is for finite
values of $U$ finite. Such a value depends on the minimum
excitation-energy value for the creation of a $-1/2$ holon. In the
case of the Hamiltonian (\ref{H}), this minimum energy value is
for electronic densities $n$ such that $0\leq n\leq 1/a$ given by
$\Delta E=E_u$. For a zero spin density $m=0$ initial ground
state, the energy $E_u=2\mu\, (U/t,n)$ equals twice the chemical
potential and is an increasing function of the on-site repulsion
$U$ with the following limiting values,

\begin{eqnarray}
E_u & = & 4t\cos (\pi na/2) \, ; \hspace{0.5cm} U/t\rightarrow 0
\,
, \nonumber \\
& = & U+ 4t\cos (\pi na)  \, ; \hspace{0.5cm} U/t\rightarrow\infty
\, . \label{EuU}
\end{eqnarray}
Moreover, this energy parameter for any value of $U/t$ is a
decreasing function of the electronic density $n$ such that,

\begin{eqnarray}
E_u & = & U + 4t \, ; \hspace{0.5cm} n\rightarrow 0 \,
, \nonumber \\
& = & E_{MH}  \, ; \hspace{0.5cm} n\rightarrow 1 \, , \label{Eun}
\end{eqnarray}
where $E_{MH}$ is the half-filling Mott-Hubbard gap \cite{Lieb}.
On the other hand, if one uses as initial state the same $m=0$
ground state but considers the excitation energy of the "simple"
Hubbard Hamiltonian (\ref{HH}), there is a shift in the definition
of the ground-state zero-energy level. For this Hamiltonian the
minimum excitation-energy value for creation of a $-1/2$ holon is
simply given by $U$.

Since rotated-electron double occupation equals the number of
$-1/2$ holons, the minimum excitation-energy value for an excited
energy eigenstate with rotated electron double occupation $M'\geq
0$ is $\Delta E=M'\,E_u$. This justifies why all gapless
excitation branches are associated with zero rotated-electron
double occupation states. The excitation energy $\Delta E=M'\,E_u$
is measured from the ground-state level and corresponds to the
Hamiltonian (\ref{H}). If instead we consider the ground-state
zero-energy level associated with the Hamiltonian (\ref{HH}), the
value of such an excitation energy becomes $M'\,U$.

Any operator $\hat{O}$ can be expanded as follows,

\begin{equation}
\hat{O} = \sum_{M\in Z}{\hat{O}}_{ME_u} \, , \label{M}
\end{equation}
where $M =...-3, -2, -1, 0, 1, 2, 3, ...$ is such that application
of the operator ${\hat{O}}_{ME_u}$ onto any eigenstate of rotated
electron double occupation of eigenvalue $M'$ leads to a final
state with rotated electron double occupancy $[M'+M]\geq 0$. Also
the corresponding rotated operator $\tilde{O}$ such that,

\begin{equation}
\tilde{O} = {\hat{V}}^{\dag}(U/t)\,\hat{O}\,{\hat{V}}(U/t) \, ,
\label{Otil}
\end{equation}
can be decomposed as,

\begin{equation}
\tilde{O} = \sum_{M\in Z}{\tilde{O}}_{ME_u} \, . \label{Mtil}
\end{equation}
The index $ME_u$ of expressions (\ref{M}) and (\ref{Mtil}) plays
the same role as the index $nU$ of Ref. \cite{Eskes}. While $M=n$,
the energy $E_u$ only equals $U$ in the $U/t\rightarrow\infty$
limit. Since our study refers to all values of $U/t$, here we use
$E_u$ instead of $U$ to label the energy of the different upper
Hubbard bands defined above. For finite values of $U/t$ the choice
of Ref. \cite{Eskes} refers to the zero-energy level of the
Hamiltonian (\ref{HH}), whereas our choice corresponds to the
ground-state zero-energy level of the more suitable Hamiltonian
(\ref{H}). Our choice is justified by the fact that in real
experiments the excitation energy is, in general, measured from
the Fermi level. Our notation tells that the minimal amount of
excitation energy for a transition which results from application
of an operator component ${\hat{O}}_{ME_u}$ onto any eigenstate of
rotated electron double occupation is $ME_u$. This excitation
energy can be positive ($M>0$) or negative ($M<0$). We note that
application onto a state of rotated-electron double occupation
$M'<\vert\,M\vert$ of a $M<0$ operator component
${\hat{O}}_{ME_u}$ gives zero. Since for the present case of
electronic densities $n$ such that $0\leq n\leq 1/a$, the ground
state has zero rotated-electron double occupation, all the final
excited states of the general ${\cal{N}}$-electron spectral
function (\ref{ONsf})-(\ref{ONMsf}) have $M\geq 0$.

The operators ${\hat{V}}^{\dag}(U/t)$ and ${\hat{V}}(U/t)$
associated with the electron - rotated-electron unitary
transformation can be written as,

\begin{equation}
{\hat{V}}^{\dag}(U/t) = e^{-\hat{S}} \, ; \hspace{0.5cm}
{\hat{V}}(U/t) = e^{\hat{S}} \, . \label{SV}
\end{equation}
These operators are uniquely defined by Eqs. (44) and (45) of the
companion paper \cite{I} and can be expressed in the form of an
expansion in $t/U$ for the operator $\hat{S}$ \cite{Mac}. An
important role in such an expansion is played by the rotated
kinetic energy operator $\tilde{T}$, associated with the operator
(\ref{Top}). The decomposition (\ref{Mtil}) of this operator leads
to three terms,

\begin{equation}
\tilde{T} = {\tilde{T}}_0 + {\tilde{T}}_{E_u} + {\tilde{T}}_{-E_u}
\, , \label{Ttildec}
\end{equation}
where

\begin{eqnarray}
{\tilde{T}}_0 & = & -t\sum_{j =1}^{N_a}\sum_{\sigma =\uparrow
,\downarrow}\sum_{\delta
=-1,+1}\Bigl[(1-{\tilde{n}}_{j,\,\bar{\sigma}})\,
{\tilde{c}}_{j,\,\sigma}^{\dag}\,{\tilde{c}}_{j+\delta,\,\sigma}\,
(1-{\tilde{n}}_{j+\delta,\,\bar{\sigma}}) \nonumber \\
& + & {\tilde{n}}_{j,\,\bar{\sigma}}\,
{\tilde{c}}_{j,\,\sigma}^{\dag}\,{\tilde{c}}_{j+\delta,\,\sigma}\,
{\tilde{n}}_{j+\delta,\,\bar{\sigma}}\Bigr] \, , \label{Ttil0}
\end{eqnarray}

\begin{equation}
{\tilde{T}}_{E_u} = -t\sum_{j =1}^{N_a}\sum_{\sigma =\uparrow
,\downarrow}\sum_{\delta =-1,+1}{\tilde{n}}_{j,\,\bar{\sigma}}\,
{\tilde{c}}_{j,\,\sigma}^{\dag}\,{\tilde{c}}_{j+\delta,\,\sigma}\,
(1-{\tilde{n}}_{j+\delta,\,\bar{\sigma}}) \, , \label{Ttil+1}
\end{equation}
and

\begin{equation}
{\tilde{T}}_{-E_u} = -t\sum_{j =1}^{N_a}\sum_{\sigma =\uparrow
,\downarrow}\sum_{\delta
=-1,+1}(1-{\tilde{n}}_{j,\,\bar{\sigma}})\,
{\tilde{c}}_{j,\,\sigma}^{\dag}\,{\tilde{c}}_{j+\delta,\,\sigma}\,
{\tilde{n}}_{j+\delta,\,\bar{\sigma}} \, . \label{Ttil-1}
\end{equation}
In these expressions $\bar{\sigma}$ is such that
$\bar{\uparrow}=\downarrow$ and $\bar{\downarrow}=\uparrow$. The
above-mentioned expansion of the operator $\hat{S}$ of the
expressions of Eq. (\ref{SV}) in powers of $t/U$ is obtained by
systematic use of Eqs. (44) and (45) of the companion paper
\cite{I}, with the result,

\begin{equation}
\hat{S} = {1\over U}\,\Bigl({\tilde{T}}_{E_u} -
{\tilde{T}}_{-E_u}\Bigr) + {1\over U^2}\,\Bigl[{\tilde{T}}_{E_u} +
{\tilde{T}}_{-E_u},\,{\tilde{T}}_0\Bigr] + ... \, , \label{Sexp}
\end{equation}
where the components of the rotated kinetic energy operator are
given in Eqs. (\ref{Ttil0})-(\ref{Ttil-1}). While equations (44)
and (45) of Ref. \cite{I} uniquely define the electron -
rotated-electron unitary operator ${\hat{V}}(U/t)$ of Eq.
(\ref{SV}) for all values of $U/t$, it is difficult to extract a
closed form expression for that operator from these equations.

Let us consider the case of few-electron operators
${\hat{O}}_{{\cal{N}}}(k)$ whose spectral-function excitation
energy is usually considered negative. Example of such operators
are the one-electron spin $\sigma$ annihilation operator of
momentum $k$ and the two-electron singlet or triplet Cooper pair
annihilation operator of momentum $k$. We note that according to
the results of the companion paper \cite{I}, the number of
rotated-electron empty sites equals the number $M_{c,\,+1/2}$ of
$+1/2$ holons. According to Eq. (C24) of the same paper, for
values of the electronic density $n$ such that $0\leq n\leq 1/a$
the ground state has a number of $+1/2$ holons whose value is
given by $M_{c,\,+1/2}^0=[N_a-N]$. Since the number of
rotated-electron empty sites is a good quantum number, we can
write a general ${\cal{N}}$-electron spectral function associated
with such a type of few-electron operators as follows,

\begin{equation}
B_{\cal{N}}(k,\,\omega) = \sum_{{\bar{M}}=0,\pm1,\pm 2,...
}^{\pm\infty}\,B_{{\cal{N}},{\bar{M}}}(k,\,\omega) \, ,
\label{ONsfB}
\end{equation}
where

\begin{equation}
B_{{\cal{N}},{\bar{M}}}(k,\,\omega) =\sum_{j}\, \vert\langle
{\bar{M}},j\vert {\hat{O}}_{{\cal{N}}}(k) \vert
GS\rangle\vert^2\,\delta\Bigl( \omega +\omega_{{\bar{M}},j}\Bigr)
\, , \label{ONMsfB}
\end{equation}
the excitation energy $\omega$ is negative, the ${\bar{M}}\equiv [
M_{c,\,+1/2}-M_{c,\,+1/2}^0]$ summation of Eq. (\ref{ONsfB}) runs
over the values of the deviation in the number of rotated-electron
empty sites ${\bar{M}}=0,\pm 1,\pm 2,...,\pm\infty$ of the excited
energy eigenstates $\vert {\bar{M}},j\rangle$, the $j$ summation
of Eq. (\ref{ONMsfB}) runs over all available excited energy
eigenstates $\vert {\bar{M}},j\rangle$ with the same value
$M_{c,\,+1/2}=[M_{c,\,+1/2}^0+{\bar{M}}]$ for the number of
rotated-electron empty sites, and $\omega_{{\bar{M}},j}\equiv
[E_{{\bar{M}},j}-E_{GS}]$ are the excitation energies relative to
the initial ground state. Note that in the case of half filling
the ground-state value for the number of rotated-electron empty
sites is zero and the ${\bar{M}}$ summation of Eq. (\ref{ONsfB})
runs over positive values of rotated-electron empty sites
${\bar{M}}=0,1,2,...,\infty$ only. Moreover, for electronic
densities $n>1/a$ the role of the numbers of rotated-electron
doubly occupied and empty sites is interchanged. For these
densities the summations over the values of rotated-electron
doubly occupied and empty sites of Eqs. (\ref{ONsf}) and
(\ref{ONsfB}), respectively, correspond to both positive and
negative integers and to positive integers, respectively. This is
because for such densities the ground state has no $+1/2$ holons.
In this paper we consider mostly the case of electronic densities
$n<1/a$, where the ground state has no $-1/2$ holons.

As further discussed in Sec. V, the first step for the evaluation
of the general few-electron spectral functions
(\ref{ONsf})-(\ref{ONMsf}) and (\ref{ONsfB})-(\ref{ONMsfB}) is the
expression of the ${\cal{N}}$-electron operator
${\hat{O}}_{\cal{N}}(k)$ in terms of the corresponding rotated
operator ${\tilde{O}}_{\cal{N}}(k)$ defined in Eq. (\ref{Otil}) as
follows \cite{V},

\begin{eqnarray}
{\hat{O}}_{\cal{N}}(k) & = &
{\hat{V}}(U/t)\,{\tilde{O}}_{\cal{N}}(k)\,{\hat{V}}^{\dag}(U/t)
\nonumber \\
& = & {\tilde{O}}_{\cal{N}}(k) +
[{\hat{S}},\,{\tilde{O}}_{\cal{N}}(k)\,] + {1\over
2}\,[{\hat{S}},\,[{\hat{S}},\,{\hat{O}}_{\cal{N}}(k)\,]\,] + ...
\, . \label{ONtil}
\end{eqnarray}
According to the results of the companion papers \cite{I,III}, the
relation of the rotated operators on the right-hand side of Eq.
(\ref{ONtil}) to the quantum objects whose occupancy
configurations describe the ground state and excited states $\vert
M,j\rangle$ and $\vert {\bar{M}},j\rangle$ of the few-electron
spectral functions (\ref{ONsf})-(\ref{ONMsf}) and
(\ref{ONsfB})-(\ref{ONMsfB}), respectively, is well defined.
Expression of the second expression of Eq. (\ref{ONtil}) in terms
of the creation and annihilation operators for these quantum
objects plays a crucial role in the evaluation of the matrix
elements of the few-electron spectral functions (\ref{ONMsf}) and
(\ref{ONMsfB}). This justifies the importance of the holon,
spinon, and $c$ pseudoparticle description introduced in the
companion paper \cite{I} and further studied here and in the
companion paper \cite{III} for the study of the few-electron
spectral properties at finite values of the excitation energy.
Solution of this problem is a direct application of the relation
of rotated electrons to the quantum numbers of the Bethe-ansatz
solution and $\eta$-spin and spin symmetries found in the
companion paper \cite{I}. Such a breakthrough provided the
relation of electrons and rotated electrons to the above-mentioned
quantum objects.

In the ensuing section we find that the first operator term of the
second expression on the right-hand side of Eq. (\ref{ONtil})
corresponds to the dominant microscopic physical processes that
control the few-electron spectral-weight properties. In the case
of ${\cal{N}}=1$-electron operators we find that more than 99\% of
the electronic spectral weight generated by application of the
operator ${\hat{O}}_{\cal{N}}$ onto a ground state corresponds to
application of the corresponding ${\cal{N}}$-rotated-electron
operator ${\tilde{O}}_{\cal{N}}$ onto the same state. This latter
operator has a simple expression in terms of holon and spinon
elementary operators, and it is associated with the dominant holon
and spinon microscopic mechanisms that control the
${\cal{N}}$-electron spectral properties.

\subsection{EXACT HOLON, SPINON, AND $c$ PSEUDOPARTICLE SELECTION RULES FOR
ROTATED-ELECTRON OPERATORS}

Let $\vert\phi\rangle$ be any eigenstate of the spin $\sigma$
electron number operator (\ref{Nsi}) and ${\hat{O}}_{{\cal{N}}}$
(or ${\tilde{O}}_{{\cal{N}}}$) a general ${{\cal{N}}}$-electron
(or ${{\cal{N}}}$-rotated-electron) operator. The corresponding
state ${\hat{O}}_{{\cal{N}}}\vert\phi\rangle$ (or
${\tilde{O}}_{{\cal{N}}}\vert\phi\rangle$) is also an eigenstate
of the spin $\sigma$ electron number operator (\ref{Nsi}). Such a
state belongs to an electron ensemble space whose electron numbers
differ from the numbers of the initial state by deviations $\Delta
N_{\uparrow}$ and $\Delta N_{\downarrow}$, such that $\Delta N =
\Delta N_{\uparrow} +\Delta N_{\downarrow}$. Since the number of
spin $\sigma$ electrons equals the number of spin $\sigma$ rotated
electrons, these deviations also refer to the latter numbers.
These deviations can be expressed in terms of the numbers
${\cal{N}}_{l_c,\,l_s}$ of Eq. (\ref{d}) as follows,

\begin{equation}
\Delta N_{\uparrow} = \sum_{l_c=\pm 1}(-l_c)\,{\cal{N}}_{l_c,\,+1}
\, ; \hspace{1cm} \Delta N_{\downarrow} = \sum_{l_c=\pm
1}(-l_c)\,{\cal{N}}_{l_c,\,-1} \, . \label{Nupdodll}
\end{equation}

Following the results of Refs. \cite{I,II}, let us consider the
four expectation values $R_{\alpha,\,l_{\alpha}}=\langle
{\hat{R}}_{\alpha,\,l_{\alpha}}\rangle$, where $\alpha =c,s$ and
$l_{\alpha}=-1,\,+1$ and the corresponding operator
${\hat{R}}_{c,\,-1}$ counts the number of electron doubly-occupied
sites, ${\hat{R}}_{c,\,+1}$ counts the number of electron empty
sites, ${\hat{R}}_{s,\,-1}$ counts the number of spin-down
electron singly-occupied sites, and ${\hat{R}}_{s,\,+1}$ counts
the number of spin-up electron singly-occupied sites. These
operators are given in Eqs. (23) and (24) of the companion paper
\cite{I}. In reference \cite{II} it was found that in the limit
$U/t\rightarrow\infty$ the deviations $\Delta D\equiv \Delta
R_{c,\,-1}$, $\Delta R_{c,\,+1}$, $\Delta R_{s,\,-1}$, and $\Delta
R_{s,\,+1}$ generated by application of a general
${\cal{N}}$-electron operator ${\hat{O}}_{\cal{N}}$ onto any
eigenstate of the spin $\sigma$ electron number operator
(\ref{Nsi}) are restricted to the ranges of the inequalities
(60)-(63) of the same reference. Obviously, given one of the four
ranges of values defined by these four inequalities, the other
three are dependent and follow from the relations given in Eq.
(56) of Ref. \cite{II}.

Let us consider the deviations $\Delta R_{c,\,-1}$, $\Delta
R_{c,\,+1}$, $\Delta R_{s,-1}$, and $\Delta R_{s,+1}$ in the
expectation values of the four operators given in Eqs. (23) and
(24) of Ref. \cite{I} in the case that both the corresponding
initial state $\vert\phi_{l}\rangle$ and final state
$\vert\phi_{l'}\rangle$ are energy eigenstates  of the 1D Hubbard
model in the limit $U/t\rightarrow\infty$. Note that it is assumed
that the final energy eigenstate $\vert\phi_{l'}\rangle$ is
contained in the state ${\hat{O}}_{\cal{N}}\vert\phi_l\rangle$
where ${\hat{O}}_{\cal{N}}$ is the general ${\cal{N}}$-electron
operator under consideration. In the case of these initial and
final states the above-mentioned deviations are such that,

\begin{eqnarray}
\Delta R_{\alpha,\,l_{\alpha}} & = &
\langle\phi_{l'}\vert{\hat{R}}_{\alpha,\,l_{\alpha}}
\vert\phi_{l'}\rangle -
\langle\phi_l\vert{\hat{R}}_{\alpha,\,l_{\alpha}}
\vert\phi_l\rangle \, ; \nonumber \\
& = &
\langle\psi_{l'}(U/t)\vert\,{\hat{V}}^{\dag}(U/t)\,{\hat{R}}_{\alpha,\,l_{\alpha}}\,
{\hat{V}}(U/t) \vert\psi_{l'}(U/t)\rangle \nonumber \\
& - &
\langle\psi_{l}(U/t)\vert\,{\hat{V}}^{\dag}(U/t)\,{\hat{R}}_{\alpha,\,l_{\alpha}}\,
{\hat{V}}(U/t) \vert\psi_{l}(U/t)\rangle \, ; \nonumber \\
& = &
\langle\psi_{l'}(U/t)\vert{\hat{M}}_{\alpha,\,\sigma_{\alpha}}
\vert\psi_{l'}(U/t)\rangle -
\langle\psi_l(U/t)\vert{\hat{M}}_{\alpha,\,\sigma_{\alpha}}
\vert\psi_l(U/t)\rangle \, , \label{DRDMee}
\end{eqnarray}
for all values of $U/t$. In order to derive this result we used
Eq. (52) of the companion paper \cite{I}. Also Eqs. (54)-(57) of
the same paper were used in the evaluation of the equalities on
the right-hand side of Eq. (\ref{DRDMee}). Equation (52) of Ref.
\cite{I} relates the energy eigenstates of the
$U/t\rightarrow\infty$ Hubbard model to the corresponding energy
eigenstates of the Hubbard model for any value of $U/t$. We note
that the last quantity on the right-hand side of Eq.
(\ref{DRDMee}) is nothing but the deviation

\begin{equation}
\Delta M_{\alpha,\,\sigma_{\alpha}} =
\langle\psi_{l'}(U/t)\vert{\hat{M}}_{\alpha,\,\sigma_{\alpha}}
\vert\psi_{l'}(U/t)\rangle -
\langle\psi_l(U/t)\vert{\hat{M}}_{\alpha,\,\sigma_{\alpha}}
\vert\psi_l(U/t)\rangle = M_{\alpha,\,\sigma_{\alpha}}' -
M_{\alpha,\,\sigma_{\alpha}} \, , \label{DMasee}
\end{equation}
in the particular case when the initial state
$\vert\psi_{l}(U/t)\rangle$ and final state
$\vert\psi_{l'}(U/t)\rangle$ are energy eigenstates of the
finite-$U/t$ Hubbard model. The final energy eigenstate
$\vert\psi_{l'}(U/t)\rangle$ is contained in the state
${\tilde{O}}_{\cal{N}}\vert\psi_{l}(U/t)\rangle$, where the
rotated operator ${\tilde{O}}_{\cal{N}}$ is related to the
operator ${\hat{O}}_{\cal{N}}$ by Eq. (\ref{ONtil}). The operator

\begin{equation}
[{\hat{O}}_{\cal{N}} - {\tilde{O}}_{\cal{N}}] =
[{\hat{S}},\,{\tilde{O}}_{\cal{N}}\,] + {1\over
2}\,[{\hat{S}},\,[{\hat{S}},\,{\hat{O}}_{\cal{N}}\,]\,] + ... \, ,
\label{DONtil}
\end{equation}
can be expressed as a sum of ${\cal{N'}}$-electron operators such
that ${\cal{N'}}>{\cal{N}}$. Note that Eq. (\ref{DONtil}) contains
the same information as Eq. (\ref{ONtil}).

The quantities $M_{\alpha,\,\sigma_{\alpha}}$ and
$M_{\alpha,\,\sigma_{\alpha}}'$ of Eq. (\ref{DMasee}) are the
eigenvalues of the energy eigenstates $\vert\psi_{l}(U/t)\rangle$
and $\vert\psi_{l'}(U/t)\rangle$, respectively, relative to the
holon ($\alpha =c$) or spinon ($\alpha =s$) number operator
${\hat{M}}_{\alpha,\,\sigma_{\alpha}}$ studied in the companion
paper \cite{I}. These two states are related to the above initial
state $\vert\phi_{l}\rangle$ and final state
$\vert\phi_{l'}\rangle$ by the transformation given in Eq. (52) of
the same paper. The equalities of Eq. (\ref{DRDMee}) are valid for
all possible $4^{N_a}$ initial energy eigenstates
$\vert\phi_l\rangle$ of the 1D Hubbard model in the limit
$U/t\rightarrow\infty$ such that $l=1,2,...,4^{N_a}$ and for final
energy eigenstates $\vert\phi_{l'}\rangle$ which are contained in
${\hat{O}}_{\cal{N}}\vert\phi_l\rangle$. According to these
equalities, the deviations in the eigenvalues of the four
operators ${\hat{R}}_{\alpha,\,l_{\alpha}}$ of Eqs. (23) and (24)
of the companion paper \cite{I} which count the number of electron
doubly-occupied sites, empty sites, spin-down singly-occupied
sites, and spin-up singly-occupied sites equal the deviations in
the eigenvalues of the four operators
${\hat{M}}_{\alpha,\,\sigma_{\alpha}}$ given in Eq. (22) of the
same paper. These four operators count the number of $-1/2$
holons, $+1/2$ holons, $-1/2$ spinons, and $+1/2$ spinons,
respectively. Note that according to Eqs. (52) and (54)-(57) of
Ref. \cite{I} such an equality is valid because the initial and
final states of the latter deviations are the energy eigenstates
of the finite-$U/t$ 1D Hubbard model $\vert\psi_l(U/t)\rangle =
\hat{V}^{\dag}(U/t)\vert\phi_l\rangle$ and
$\vert\psi_{l'}(U/t)\rangle =
\hat{V}^{\dag}(U/t)\vert\phi_{l'}\rangle$, respectively. Here the
final state $\vert\psi_{l'}(U/t)\rangle =
\hat{V}^{\dag}(U/t)\vert\phi_{l'}\rangle$ is contained in the
state ${\tilde{O}}_{\cal{N}}\vert\psi_{l}(U/t)\rangle=
\hat{V}^{\dag}(U/t)\,{\hat{O}}_{\cal{N}}\vert\phi_l\rangle$.

The electron double-occupation, no-occupation, spin-down
single-occupation, and spin-up single-occupation eigenvalue
deviations (\ref{DRDMee}) which result from application onto the
state $\vert\phi_l\rangle$ of the above general operator
${\hat{O}}_{\cal{N}}$ are restricted to the ranges of the
inequalities (60)-(63) of Ref. \cite{II}. Therefore, also the the
$-1/2$-holon-number, $+1/2$-holon-number, $-1/2$-spinon-number,
and $+1/2$-spinon-number eigenvalue deviations (\ref{DMasee})
which result from application onto the state
$\vert\psi_l(U/t)\rangle$ of the corresponding
${\cal{N}}$-rotated-electron operator ${\tilde{O}}_{\cal{N}}$ are
restricted to the ranges of the following inequalities,

\begin{equation}
-\sum_{l_s=\pm 1} {\cal{N}}_{+1,\,l_s}\leq\Delta
M_{c,\,-1/2}\leq\sum_{l_s=\pm 1} {\cal{N}}_{-1,\,l_s} \, ,
\label{Mc-range}
\end{equation}

\begin{equation}
-\sum_{l_s=\pm 1} {\cal{N}}_{-1,\,l_s}\leq\Delta
M_{c,\,+1/2}\leq\sum_{l_s=\pm 1} {\cal{N}}_{+1,\,l_s} \, ,
\label{Mc+range}
\end{equation}

\begin{equation}
-\sum_{l_c,l_s=\pm 1}\delta_{l_c,\,-l_s}\,
{\cal{N}}_{l_c,\,l_s}\leq\Delta M_{s,\,-1/2}\leq\sum_{l_c,l_s=\pm
1}\delta_{l_c,\,l_s}\, {\cal{N}}_{l_c,\,l_s} \, , \label{Ms-range}
\end{equation}
and

\begin{equation}
-\sum_{l_c,l_s=\pm 1}\delta_{l_c,\,l_s}\,
{\cal{N}}_{l_c,\,l_s}\leq\Delta M_{s,\,+1/2}\leq\sum_{l_c,l_s=\pm
1}\delta_{l_c,\,-l_s}\, {\cal{N}}_{l_c,\,l_s} \, .
\label{Ms+range}
\end{equation}
According to Eq. (\ref{DMasee}), the eigenvalues
$M_{\alpha,\,\sigma_{\alpha}}$ and $M_{\alpha,\,\sigma_{\alpha}}'$
of these deviations refer to an initial state which is one of the
$4^{N_a}$ energy eigenstates $\vert\psi_l (U/t)\rangle$ of the
finite-$U/t$ Hubbard model and to a final state which is an energy
eigenstate $\vert\psi_{l'}(U/t)\rangle$ of the same model
contained in the state
${\tilde{O}}_{\cal{N}}\vert\psi_l(U/t)\rangle$. However, since the
$4^{N_a}$ states $\vert\psi_l(U/t)\rangle$ associated with the
deviations of Eq. (\ref{DMasee}) constitute a complete basis for
the Hilbert space of the 1D Hubbard model, the inequalities
(\ref{Mc-range})-(\ref{Ms+range}) also apply to general deviations
of the form,

\begin{equation}
\Delta M_{\alpha,\,\sigma_{\alpha}} =
\langle\psi'\vert{\hat{M}}_{\alpha,\,\sigma_{\alpha}}
\vert\psi'\rangle -
\langle\psi\vert{\hat{M}}_{\alpha,\,\sigma_{\alpha}}
\vert\psi\rangle  \, , \label{DMall}
\end{equation}
provided that the arbitrary states $\vert\psi'\rangle$ and
$\vert\psi\rangle$ are eigenstates of the spin $\sigma$ electron
number operator (\ref{Nsi}).

It follows from the relations of Eq. (50) of the companion paper
\cite{I} that out of the four inequalities
(\ref{Mc-range})-(\ref{Ms+range}), only one is independent. By
summation of the two inequalities (\ref{Mc-range}) and
(\ref{Mc+range}) and of the two inequalities (\ref{Ms-range}) and
(\ref{Ms+range}) we find the following inequalities for the
deviations in the total number of holons and spinons,

\begin{equation}
-{\cal{N}}\leq\Delta M_c\leq {\cal{N}} \, ; \hspace{1cm}
-{\cal{N}}\leq\Delta M_s\leq {\cal{N}} \, . \label{DMcsrange}
\end{equation}
Note that the limiting values of the inequalities
(\ref{DMcsrange}) are simply $-\cal{N}$ and $\cal{N}$, where
$\cal{N}$ is the number of elementary rotated-electron operators
of the expression of the operator ${\tilde{O}}_{\cal{N}}$.
Finally, through the use of Eq. (28) of Ref. \cite{I} we find that
the validity of these inequalities is equivalent to the
inequalities

\begin{equation}
-{\cal{N}}\leq\Delta N_c\leq {\cal{N}} \, ; \hspace{0.5cm}
-{\cal{N}}\leq\Delta N_c^h\leq {\cal{N}} \, , \label{DNc}
\end{equation}
for the deviations $\Delta N_c=-\Delta N_c^h$ in the number $ N_c$
of $c$ pseudoparticles and in the number $N_c^h$ of $c$
pseudoparticle holes, respectively.

We emphasize that the exact holon and spinon selection rules
(\ref{Mc-range})-(\ref{Ms+range}) and (\ref{DMcsrange}) refer to
deviations generated by ${\cal{N}}$-rotated-electron operators. We
find below that in the case of deviations generated by the
corresponding ${\cal{N}}$-electron operator the dominant holon and
spinon microscopic physical processes lead to excited states that
obey the inequalities (\ref{Mc-range})-(\ref{Ms+range}) and
(\ref{DMcsrange}). Such dominant microscopic physical processes
control most of the ${\cal{N}}$-electron spectral-weight.
Interestingly, as in the $U>>t$ case, for $U<<t$ these rules are
exact for the corresponding ${\cal{N}}$-electron operator, the
dominant processes becoming the only processes contributing to the
${\cal{N}}$-electron spectral properties. On the other hand, for
values of the on-site repulsion such that $U\approx 4t$ these
rules are not exact for ${\cal{N}}$-electron operators but
correspond to dominant processes that amount for more than 99\% of
the electronic spectral weight. For instance, in the case of
one-electron creation operators, we find that the spectral weight
generated by higher-order holon and spinon processes is extremely
small. Such a weight corresponds to states with holon and spinon
numbers outside the domains defined by inequalities
(\ref{Mc-range})-(\ref{Ms+range}) and (\ref{DMcsrange}) and is
maximum for half filling and at $U\approx 4t$ where it amounts for
about 0.25\% of the total spectral weight. However, such a weight
decreases for decreasing values of the electronic density. For
instance, at quarter filling it is maximum also for $U\approx 4t$
but amounts for 0.005\% of the total spectral weight only.
Moreover, for small electronic densities the dominant processes
become the only processes for ${\cal{N}}$-electron operators at
any value of $U/t$. In the case of the one-electron creation
operators the states generated by the dominant processes refer to
a {\it minimum} amount of relative spectral weight for $U\approx
4t$. Nevertheless, such a {\it minimum} value corresponds for the
dominant holon and spinon processes to 99.75\%, 99.99\%, and
100.00\% of the total one-electron addition spectral weight for
electronic densities $n=1/a$, $n=1/2a$, and $n<<1/a$,
respectively, as found in Sec. IV.

\subsection{THE EFFECTIVE ELECTRONIC LATTICE}

The electron - rotated-electron unitary transformation performs a
rotation in Hilbert space which maps electrons onto
rotated-electrons such that rotated-electron double occupation is
a good quantum number. As discussed below, the rotated-electron
site distribution configurations that describe the energy
eigenstates are independent on the value of the ratio $U/t$. In
contrast, the electron site distribution configurations that
describe these states are dependent on $U$. This is consistent
with the $U/t$ dependence of the electron double-occupation
quantities studied in Ref. \cite{II}.

As discussed in the companion paper \cite{I}, the electronic
lattice remains invariant under the electron - rotated-electron
unitary transformation. However, in order to distinguish the
rotated-electron from the electronic site distribution
configurations it is useful to introduce an {\it effective
electronic lattice}. The rotated electrons occupy the sites of the
effective electronic lattice, whereas the electrons occupy the
sites of the {\it real-space lattice}. Such an effective
electronic lattice has the same number of sites $j=1,2,3,...,N_a$,
lattice constant $a$, and length $L=N_a\times a$ as the real-space
lattice.

In the companion paper \cite{III} it is found that the
rotated-electron site distribution configurations that describe
the energy eigenstates include separated charge and spin
sequences. The charge (and spin) sequences correspond to
distribution configurations of the rotated-electron doubly
occupied sites and empty sites (and spin-down and spin-up singly
occupied sites). For the effective electronic lattice the number
of $-1/2$ holons equals rotated-electron double occupation and
thus plays the same role as electron double occupation in the
real-space lattice. For finite values of $U/t$, the number of
electron doubly-occupied sites, empty sites, spin-down
singly-occupied sites, and spin-up singly-occupied sites of the
real-space lattice electron site distribution configurations that
describe the energy eigenstates, is not a good quantum number. In
contrast, for the effective electronic lattice the number of
rotated-electron doubly-occupied sites, empty sites, spin-down
singly-occupied sites, and spin-up singly-occupied sites is a good
quantum number for all values of $U/t$. This number equals the
number of $-1/2$ holons, $+1/2$ holons, $-1/2$ spinons, and $+1/2$
spinons, respectively, of any energy eigenstate.

A property of major importance is that the effective electronic
lattice site distribution configurations of the rotated electrons
which describe the energy eigenstate $\vert\psi_l(U/t)\rangle$ of
the relation (52)of the companion paper \cite{I} are the same as
the corresponding real-space lattice site distribution
configurations of the electrons which describe the state
$\vert\phi_l\rangle$. Since this latter state is an energy
eigenstate of the 1D Hubbard model in the limit
$U/t\rightarrow\infty$, the rotated-electron site distribution
configurations which describe the energy eigenstates are
independent of the value of $U/t$ and are precisely the same as
the corresponding electron site distribution configurations of the
$U/t\rightarrow\infty$ 1D Hubbard model. In the companion paper
\cite{III} it is shown that this property is behind the fact that
the $c$ and $\alpha,\nu$ pseudoparticle band-momentum occupancy
configurations which describe the energy eigenstates are also
independent of the value of $U/t$, only the energy spectrum of
such configurations are $U/t$ dependent. The fact that the
$U/t\rightarrow\infty$ selection rule ranges given in Eqs.
(60)-(63) of Ref. \cite{II} are precisely the same as the
finite-$U/t$ ranges imposed by the inequalities
(\ref{Mc-range})-(\ref{Ms+range}) results from the $U/t$
independence of the rotated-electron site distribution
configurations which describe the energy eigenstates.

Another important property is that for electronic densities
smaller or equal to one, a ground state has zero rotated-electron
double occupation. The same occurs for the ground state electron
site distribution configurations of the real-space lattice in the
limit $U/t\rightarrow\infty$, the ground-state electron double
occupation being in that limit given by $D^0=R^0_{c,\,-1}=0$. This
property is consistent with the above property that the
rotated-electron site distribution configurations which describe
the energy eigenstates are $U/t$ independent. Note that creation
of a rotated electron onto the ground state either transforms an
empty site of the effective electronic lattice into a
singly-occupied site or a singly-occupied site of the lattice into
a doubly-occupied site. These two alternative transitions lead to
$-1/2$ holon number deviations such that $\Delta M_{c,\,-1/2}=0$
and $\Delta M_{c,\,-1/2}=1$, respectively. Therefore, the
selection rule of Eq. (\ref{src}) leads to permitted final states
such that the values of these deviations are $\Delta
M_{c,\,-1/2}=0,1$. For the effective electronic lattice the
operators ${\hat{M}}_{c,-1/2}$, ${\hat{M}}_{c,+1/2}$,
${\hat{M}}_{s,-1/2}$, and ${\hat{M}}_{s,+1/2}$ count the number of
rotated-electron doubly-occupied sites, empty sites, spin-down
singly-occupied sites, and spin-up singly-occupied sites,
respectively. Each site of the effective electronic lattice can
either be doubly occupied by two rotated electrons of opposite
spin projection, empty, singly occupied by a spin-down rotated
electron, and singly occupied by an spin-up rotated electron. The
result of Ref. \cite{I} that the $-1/2$ holons are zero
spin-singlet combinations of two electrons of opposite spin
projection whereas the $+1/2$ holons are two-electronic hole
quantum objects, is consistent with the $-1/2$ holons and the
$+1/2$ holons corresponding to rotated-electron doubly-occupied
and empty sites of the effective electronic lattice, respectively.

In the companion paper \cite{III} it is found that the effective
electronic lattice introduced here is related to a set of
effective pseudoparticle lattices. The concept of effective
pseudoparticle lattice introduced in that reference arises from a
local description of the pseudoparticles, alternative to the
band-momentum $q$ description extracted directly from the
Bethe-ansatz solution in the companion paper \cite{I}. While in
the limit $U/t\rightarrow\infty$ such a local description of the
pseudoparticles refers to the real-space lattice, in the case of
finite values of $U/t$ the local pseudoparticle character refers
to the effective electronic lattice, as discussed in Ref.
\cite{III}. In that reference the local pseudoparticles are
described in terms of rotated-electron site distribution
configurations. The concepts of a effective pseudoparticle
elattice and of a local pseudoparticle introduced in the companion
paper \cite{III} are used in the studies of the few-electron
spectral properties of the model presented in Refs. \cite{IIIb,V}.

\section{THE DOMINANT HOLON AND SPINON MICROSCOPIC PHYSICAL PROCESSES FOR THE
FEW-ELECTRON SPECTRAL PROPERTIES}

The commutation relations (60)-(62) of the companion paper
\cite{I} imply that the six generators of the $\eta$-spin and spin
$SU(2)$ algebras given in Eqs. (2), (9), and (10) of the same
paper and the momentum operator (\ref{Popel}) have the same
expressions both in terms of elementary electronic operators
$c_{j,\,\sigma}^{\dag}$ and $c_{j,\,\sigma}$ and elementary
rotated-electron operators ${\tilde{c}}_{j,\,\sigma}^{\dag}$ and
${\tilde{c}}_{j,\,\sigma}$, respectively. Thus all these seven
operators are both two-electron and two-rotated-electron
operators. These two-electron operators are such that all operator
terms of the second expression of Eq. (\ref{ONtil}) vanish except
the first term. Thus the operator (\ref{DONtil}) associated with
such operators vanishes. This result suggests that application of
an operator of the form (\ref{DONtil}) associated with a general
$\cal{N}$-electron operator onto a ground state leads to very
little electronic spectral weight. The latter spectral weight
vanishes in the particular case that the ${\cal{N}}$-electron
operator commutes with the electron - rotated-electron unitary
operator, but is expected to be extremely small otherwise, at
least for few-electron operators. The reason is that such general
${\cal{N}}$-electron and ${\cal{N}}$-rotated-electron operators
are products of the same elementary operators
$c_{j,\,\sigma}^{\dag},\,c_{j,\,\sigma}$ and
${\tilde{c}}_{j,\,\sigma}^{\dag},\,{\tilde{c}}_{j,\,\sigma}$,
respectively, as the above two-electron operators. Application of
any of the above seven two-electron operators and of the
corresponding two-rotated-electron operators onto a ground state
leads to the same final states. Other two-electron or few-electron
operators have slightly different expressions in terms of the same
elementary electronic operators. Thus it is expected that the
application of any other few-electron operator onto a ground state
leads to almost the same final states as application of the
corresponding few-rotated-electron operator on the same ground
state. This is equivalent to say that application of the operator
(\ref{DONtil}) associated with such operators onto a ground state
leads to almost no spectral weight. This property is a direct
result of the unitary and canonical character of the electron -
rotated-electron transformation. This prediction is confirmed by
the quantitative results obtained below. Indeed, we find that the
dominant microscopic physical processes that control the
${\cal{N}}$-electron spectral-weight distribution of the
excitation ${\hat{O}}_{\cal{N}}\vert GS\rangle$ are associated
with the transformation laws of the ${\cal{N}}$-rotated-electron
operator of the second expression of Eq. (\ref{ONtil}). Indeed, in
the case ${\cal{N}}=1$, over 99\% percent of the spectral weight
of ${\hat{O}}_{\cal{N}}\vert GS\rangle$ is found to be contained
in ${\tilde{O}}_{\cal{N}}\vert GS\rangle$.

We start by studying the specific form that the exact holon and
spinon selection rules associated with the inequalities
(\ref{Mc-range})-(\ref{Ms+range}) take for transitions generated
by application onto a ground state of a general
$\cal{N}$-rotated-electron operator. While these ground-state
selection rules refer to ${\cal{N}}$-rotated-electron operators,
spectral functions of physical interest are of the form
(\ref{ONsf})-(\ref{ONMsf}) and (\ref{ONsfB})-(\ref{ONMsfB}) and
involve ${\cal{N}}$-electron operators. Therefore, the main
subject of this section is the study and discussion of the
consequences of the ${\cal{N}}$-rotated-electron ground-state
selection rules on the ${\cal{N}}$-electron spectral-weight
properties. We find that the holon and spinon deviation number
restrictions of the inequalities (\ref{Mc-range})-(\ref{Ms+range})
correspond in the case of excitations generated by
${\cal{N}}$-electron operators to the dominant holon and spinon
microscopic physical processes that control the electronic
spectral properties.

In the quantitative numerical studies presented below we devote
particular attention to the basic ${\cal{N}}=1$ electron and
rotated-electron operators. We provide evidence that the general
results obtained for one-electron operators also apply to
${\cal{N}}=2$ and other few-electron operators.

\subsection{THE GROUND-STATE CHARGE AND SPIN SELECTION RULES FOR
ROTATED-ELECTRON OPERATORS}

The inequalities (\ref{Mc-range})-(\ref{Ms+range}) provide the
largest domains for permitted values of $\pm 1/2$ holon and $\pm
1/2$ spinon number deviations generated by application of
$\cal{N}$-rotated-electron operators ${\tilde{O}}_{{\cal{N}}}$.
According to Eq. (\ref{d}) the number $\cal{N}$ is given by
${\cal{N}}=\sum_{l_c,\,l_s=\pm 1}{\cal{N}}_{l_c,\,l_s}$, where the
numbers ${\cal{N}}_{-1,\,-1}$, ${\cal{N}}_{-1,\,+1}$,
${\cal{N}}_{+1,\,-1}$, and ${\cal{N}}_{+1,\,+1}$ are defined below
such an equation. The deviation permitted values defined by
inequalities (\ref{Mc-range})-(\ref{Ms+range}) are determined
uniquely by the values of the four numbers ${\cal{N}}_{l_c,\,l_s}$
specific to the operator ${\tilde{O}}_{{\cal{N}}}$ and do not
depend on the values of $U/t$, electronic density $n$, and spin
density $m$. Moreover, these inequalities provide the largest
domains for permitted deviation values for all values of $U/t$,
electronic densities $0\leq n\leq 1/a$ and $1/a\leq n\leq 2/a$ and
spin densities $-n\leq m\leq n$ and $-(2/a-n)\leq m\leq (2/a-n)$,
respectively. However, the domains of such a permitted values can
be smaller. Let $M^0_{\alpha,\,\pm 1/2}$ be the numbers of $\pm
1/2$ holons ($\alpha =c$) and $\pm 1/2$ spinons ($\alpha =s$) of
the initial state. For instance, in case of deviations which do
not obey the following inequalities,

\begin{eqnarray}
M^0_{c,\,-1/2} & \geq & \sum_{l_s=\pm 1}{\cal{N}}_{+1,\,l_s} \, ;
\hspace{1cm} M^0_{c,\,+1/2}\geq\sum_{l_s=\pm 1}
{\cal{N}}_{-1,\,l_s} \, ; \nonumber \\ M^0_{s,\,-1/2} & \geq &
\sum_{l_c,l_s=\pm 1}\delta_{l_c,\,-l_s}\, {\cal{N}}_{l_c,\,l_s} \,
; \hspace{1cm} M^0_{s,\,+1/2}\geq\sum_{l_c,l_s=\pm
1}\delta_{l_c,\,l_s}\, {\cal{N}}_{l_c,\,l_s} \, , \label{InHS}
\end{eqnarray}
the corresponding domains of deviation values are smaller and are
contained in those defined by inequalities
(\ref{Mc-range})-(\ref{Ms+range}).

For simplicity in this section we consider initial ground states
corresponding to electronic densities $0\leq n\leq 1/a$ and spin
densities $0\leq m\leq n$. As discussed in the companion paper
\cite{I}, one can generate from states with electronic densities
$0\leq n\leq 1/a$ and spin densities $0\leq m\leq n$ the
corresponding towers of states with electronic densities and spin
densities belonging to the above extended domains. This is reached
by repetitive application of the off-diagonal generators of the
$\eta$-spin and spin algebras given in Eqs. (9) and (10) of Ref.
\cite{I}, respectively, onto the former states.

In the particular case when the initial state is a ground state
$\vert GS\rangle$ with values of the density and spin density
belonging to the above domains, the state
${\hat{O}}_{{\cal{N}}}\vert GS\rangle$ belongs to an electron
ensemble space with electron and rotated-electron numbers,

\begin{equation}
N = N_{\uparrow} + N_{\downarrow}=  N^0 + \Delta N \, ;
\hspace{1cm} N_{\uparrow} = N^0_{\uparrow} + \Delta N_{\uparrow}
\, ; \hspace{1cm} N_{\downarrow} = N^0_{\downarrow} + \Delta
N_{\downarrow} \, . \label{GStran}
\end{equation}
Here $N^0$, $N^0_{\uparrow}$, and $N^0_{\downarrow}$ are the
ground-state initial electron-ensemble space electron and
rotated-electron numbers and the deviations are given by the
general expressions of Eq. (\ref{Nupdodll}). (We recall that the
spin $\sigma$ electron numbers remain invariant under the electron
- rotated-electron unitary transformation and thus the number of
spin $\sigma$ electrons equals the number of spin $\sigma$ rotated
electrons.)

Let us first disregard the selection rules associated with the
inequality (\ref{Mc-range}). In this case the use of Eq.
(\ref{Nupdodll}) and Eqs. (49) and (50) of the companion paper
\cite{I} reveals that the deviations leading to the set of final
states whose $c$-pseudoparticle, $-1/2$ holon, and $-1/2$ spinon
numbers are compatible with the state ${\tilde{O}}_{\cal{N}}\vert
GS\rangle$ generated by application onto the ground state of a
general ${\cal{N}}$-rotated-electron operator are such that,

\begin{equation}
\Delta M_{c,\,-1/2} = 0, 1, 2, 3, ..., N^0_{\downarrow} +
\sum_{l_c=\pm 1}(-l_c)\,{\cal{N}}_{l_c,\,-1} = 0, 1, 2, 3, ...,
N_{\downarrow}\, . \label{domMc}
\end{equation}
The corresponding possible values of the $-1/2$ spinon number
deviations $\Delta M_{s,\,-1/2}$ are determined by the value of
the $-1/2$ holon number deviation $\Delta M_{c,\,-1/2}$ and are
given by,

\begin{equation}
\Delta M_{s,\,-1/2} = \sum_{l_c=\pm 1}(-l_c)\,{\cal{N}}_{l_c,\,-1}
- \Delta M_{c,\,-1/2} = \Delta N_{\downarrow} - \Delta
M_{c,\,-1/2} \, . \label{DMsdMc}
\end{equation}

For given initial values of the spin $\sigma$ electron and
rotated-electron numbers, only one of the two deviations $\Delta
M_{c,\,-1/2}$ and $\Delta M_{s,\,-1/2}$ is independent. Thus Eq.
(\ref{domMc}) reveals that the number of different CPHS ensemble
spaces spanned by states whose electron and rotated-electron
numbers are the same as the ones of the state
${\tilde{O}}_{\cal{N}}\vert GS\rangle$ is $N^0_{\downarrow}+\Delta
N_{\downarrow}=N_{\downarrow}$, with
$N_{\downarrow}\rightarrow\infty$ in the present thermodynamic
limit. This is confirmed by Eqs. (\ref{domMc})-(\ref{DMsdMc}).

According to the expressions given in Eq. (C24) of the companion
paper \cite{I} the number of $-1/2$ holons vanishes in the case of
a ground state and thus the inequality
$M^0_{c,\,-1/2}\geq\sum_{l_s=\pm 1}{\cal{N}}_{+1,\,l_s}$ given in
Eq. (\ref{InHS}) is not met except when $\sum_{l_s=\pm
1}{\cal{N}}_{+1,\,l_s}=0$. This shortens the domain of the
deviation values relative to the largest permitted domains given
in inequalities (\ref{Mc-range})-(\ref{Ms+range}). When the
initial state is the ground state the $-1/2$-holon, $+1/2$-holon,
$-1/2$-spinon, and $+1/2$-spinon number deviations are restricted
to the ranges of the following inequalities,

\begin{equation}
0\leq\Delta M_{c,\,-1/2}\leq\sum_{l_s=\pm 1} {\cal{N}}_{-1,\,l_s}
\, , \label{gsMc-range}
\end{equation}

\begin{equation}
\sum_{l_c,l_s=\pm 1}(l_c)\,{\cal{N}}_{l_c,\,l_s}\leq\Delta
M_{c,\,+1/2}\leq\sum_{l_s=\pm 1} {\cal{N}}_{+1,\,l_s} \, ,
\label{gsMc+range}
\end{equation}

\begin{equation}
-\sum_{l_c,l_s=\pm 1}\delta_{l_c,\,-l_s}\,
{\cal{N}}_{l_c,\,l_s}\leq\Delta M_{s,\,-1/2}\leq\sum_{l_c=\pm
1}(-l_c)\, {\cal{N}}_{l_c,\,-1} \, , \label{gsMs-range}
\end{equation}
and

\begin{equation}
-\sum_{l_c,l_s=\pm 1}\delta_{l_c,\,l_s}\,
{\cal{N}}_{l_c,\,l_s}\leq\Delta M_{s,\,+1/2}\leq\sum_{l_c=\pm
1}(-l_c)\, {\cal{N}}_{l_c,\,+1}  \, . \label{gsMs+range}
\end{equation}
Thus in the case of the $-1/2$ holon deviations $\Delta
M_{c,\,-1/2}$ only the following integer values are permitted,

\begin{equation}
\Delta M_{c,\,-1/2}= \Delta L_{c,\,-1/2} + \sum_{\nu
=1}^{\infty}\nu\,\Delta N_{c,\,\nu} =0,1,2,...,\sum_{l_s=\pm
1}{\cal{N}}_{-1,\,l_s} \, . \label{src}
\end{equation}
Here we used Eq. (30) of the companion paper \cite{I} to express
the deviation $\Delta M_{c,\,-1/2}$ in terms of the deviations in
the numbers of $-1/2$ Yang holons and $c,\nu$ pseudoparticles. We
emphasize that the exact ground-state charge selection rule
(\ref{src}) also limits the value of the deviations $\Delta N_c^h$
in the number of $c$ pseudoparticle holes. Indeed such a selection
rule is equivalent to the following inequality,

\begin{equation}
\sum_{l_c,\,l_s=\pm 1}(l_c)\,{\cal{N}}_{l_c,\,l_s}\leq\Delta N_c^h
\leq {\cal{N}} \, . \label{gSc-range}
\end{equation}
Note that this inequality is a particular case  for an initial
ground state of the general $\Delta N_c^h$ inequality (\ref{DNc}).

In general the ground-state selection rule (\ref{src}) also
implies that the maximum permitted values of the deviations
$\Delta L_{c,\,-1/2}$ and $\sum_{\nu =1}^{\infty}\nu\,\Delta
N_{c,\,\nu}$ are $\mbox{max}\{\Delta L_{c,\,-1/2}\}=\sum_{l_s=\pm
1}{\cal{N}}_{-1,\,l_s}$ and $\mbox{max}\{\sum_{\nu
=1}^{\infty}\nu\,\Delta N_{c,\,\nu}\} =\sum_{l_s=\pm
1}{\cal{N}}_{-1,\,l_s}$, respectively, provided that the condition
$\mbox{max}\{(\Delta L_{c,\,-1/2} + \sum_{\nu
=1}^{\infty}\nu\,\Delta N_{c,\,\nu})\}=\sum_{l_s=\pm
1}{\cal{N}}_{-1,\,l_s}$ is respected. However, we note that the
value $\mbox{max}\{\sum_{\nu =1}^{\infty}\nu\,\Delta N_{c,\,\nu}\}
=\sum_{l_s=\pm 1}{\cal{N}}_{-1,\,l_s}$ can be reached only if the
inequalities (i) $L^0_{c,\,+1/2}+\sum_{l_s=\pm
1}{\cal{N}}_{+1,\,l_s}\geq\sum_{l_s=\pm 1}{\cal{N}}_{-1,\,l_s}$
and (ii) $M^0_{s,\,\pm 1/2}\geq {\cal{N}}_{-1,\,\mp
1}+{\cal{N}}_{+1,\,\pm 1}$ are met. Inequality (i) follows from
the requirement that generation of a $c,\nu$ pseudoparticle
involves combination of the $\nu$ new created $-1/2$ holons with a
number $\nu$ of $+1/2$ holons \cite{I}. The number
$L^0_{c,\,+1/2}$ gives the number of these $+1/2$ holons
pre-existing in the initial state and $\sum_{l_s=\pm
1}{\cal{N}}_{+1,\,l_s}$ gives the number of $+1/2$ holons created
by annihilation of rotated electrons singly occupying sites in the
initial state. Inequality (ii) results from two requirements:
First the maximum value of $\sum_{\nu =1}^{\infty}\nu\,\Delta
N_{c,\,\nu}$ is reached when each new created spin-down and/or
spin-up rotated electron combines with a spin-up and/or spin-down
rotated electron pre-existing in rotated-electron singly occupied
sites of the initial state. This requires that $M^0_{s,\,\pm 1/2}$
should be larger than or equal to ${\cal{N}}_{-1,\,\mp 1}$; Second
at least ${\cal{N}}_{+1,\,\mp 1}$ of spin $\mp 1/2$ rotated
electrons singly occupying sites of the initial state should be
annihilated and give rise to empty sites associated with the new
created $+1/2$ holons. This requires the presence of the extra
term ${\cal{N}}_{+1,\,\pm 1}$ on the right-hand side of the above
inequality (ii).

For instance, for the half-filling $2S_c^0=L^0_c=0$ ground state
the values of the $c,\nu$ pseudoparticle number deviations are
restricted by the inequality $\sum_{\nu =1}^{\infty}\nu\,\Delta
N_{c,\,\nu}\leq \mbox{min}\{\sum_{l_s=\pm 1}{\cal{N}}_{-1,\,l_s}\,
,\sum_{l_s=\pm 1}{\cal{N}}_{+1,\,l_s}\}$. However, there might
exist other selection rules which further restrict the values of
these $c,\nu$ pseudoparticle deviations. Recent numerical results
\cite{Gu} reveal that in the case of the half-filling
$2S_c^0=L^0_c=0$ initial ground state, the state generated by
application of the ${\cal{N}}=2$ charge operator contains no $c,1$
pseudoparticles. This suggests that there is another selection
rule which states that generation of $c,\nu$ pseudoparticles by
new created rotated electrons results from combination of the
$-1/2$ holons generated by creation of these rotated electrons
with pre-existing $+1/2$ Yang holons in the initial ground state.
Since for the half-filling initial ground state we have that
$2S_c^0=L^0_c=0$, the numerical results of Ref. \cite{Gu} suggest
that in the case of such an initial ground state $\sum_{\nu
=1}^{\infty}\nu\,\Delta N_{c,\,\nu}=0$ and thus Eq. (\ref{src})
simplifies to $\Delta M_{c,\,-1/2}= \Delta L_{c,\,-1/2}
=0,1,2,...,\sum_{l_s=\pm 1}{\cal{N}}_{-1,\,l_s}$. Moreover, if
such a selection rule is valid for all
${\cal{N}}$-rotated-electron operators, the maximum permitted
value of the deviation $\sum_{\nu =1}^{\infty}\nu\,\Delta
N_{c,\,\nu}$ becomes $\mbox{max}\{\sum_{\nu
=1}^{\infty}\nu\,\Delta N_{c,\,\nu}\} =\mbox{min}\{ \sum_{l_s=\pm
1}{\cal{N}}_{-1,\,l_s}\, ,L^0_{c,\,+1/2}\}$. This gives zero in
the case of the half-filling initial ground state, in agreement
with the numerical results of Ref. \cite{Gu}.

The number $\nu$ equals the length of the ideal charge Takahashi's
string excitations associated with the $c,\nu$ pseudoparticle
branch \cite{I}. We emphasize that an infinite number of the
matrix elements between the initial ground state and excited
states of the Lehmann representation of the
${\cal{N}}$-rotated-electron spectral function of the operator
${\tilde{O}}_{{\cal{N}}}$ vanishes exactly as a consequence of the
ground-state charge selection rule associated with Eq.
(\ref{src}). Moreover, we find later in this section that the
exact ${\cal{N}}$-rotated-electron rules correspond to the
dominant holon and spinon microscopic physical processes generated
by application of the corresponding ${\cal{N}}$-electron operator
onto the ground state. For finite values of $U/t$ less than 1\% of
the ${\cal{N}}$-electron spectral weight does not correspond to
the final states associated with deviation values obeying the
selection rule (\ref{src}). Such a small amount of spectral weight
corresponds to a few extra final states with deviation values
outside the ranges given in Eq. (\ref{src}). However, it is found
in Ref. \cite{V} that the absolute value of the matrix elements of
the spectral functions (\ref{ONsf})-(\ref{ONMsf}) and
(\ref{ONsfB})-(\ref{ONMsfB}) decreases rapidly as the number of
holon and spinon processes needed for generation of the
corresponding excited states increases. It follows that also for
the ${\cal{N}}$-electron operator an infinite number of possible
final states give no measurable contribution to the
${\cal{N}}$-electron spectral weight. This property is related to
the exact ground-state selection rule (\ref{src}) for the states
generated by the corresponding ${\cal{N}}$-rotated-electron
operator.

In the case of ${\cal{N}}$-rotated-electron operators, the
absolute maximum value of the permitted values given in Eq.
(\ref{src}) is given by ${\cal{N}}$. The value
$M_{c,\,-1/2}={\cal{N}}$ is reached when the operator
${\tilde{O}}_{\cal{N}}$ can be written as the product of
${\cal{N}}$ creation rotated-electron operators. Analysis of Eq.
(\ref{src}) reveals that the general ${\cal{N}}$-rotated-electron
(and ${\cal{N}}$-electron) excitation ${\tilde{O}}_{\cal{N}}\vert
GS\rangle$ has quantum overlap (and ${\hat{O}}_{\cal{N}}\vert
GS\rangle$ has significant quantum overlap) only for energy
eigenstates described by charge Takahashi's ideal string
excitations of length $\nu \leq \sum_{l_s=\pm
1}{\cal{N}}_{-1,\,l_s}$. In the particular case when
${\tilde{O}}_{\cal{N}}$ can be expressed as a product of $\cal{N}$
creation rotated-electron operators this gives $\nu \leq
{\cal{N}}$. The number ${\cal{N}}$ is thus the maximum absolute
value for the length of such a charge ideal string excitation that
can be generated by application onto the ground state of a general
${\cal{N}}$-rotated-electron operator. In the case of the
excitation generated by application onto the same state of the
corresponding ${\cal{N}}$-electron operator more than 99\% of the
spectral weight is exhausted by excited states containing string
excitations of length $\nu \leq \sum_{l_s=\pm
1}{\cal{N}}_{-1,\,l_s}$, as confirmed below for one-electron
operators. This also gives $\nu \leq {\cal{N}}$ if the
${\cal{N}}$-electron operator can be written as a product of
$\cal{N}$ creation electronic operators.

When the initial state is a ground state the permitted values of
the $-1/2$ spinon deviation $\Delta M_{s,\,-1/2}$ generated by
application of the ground state of a ${\cal{N}}$-rotated-electron
operator are given by Eq. (\ref{DMsdMc}), with $\Delta
M_{c,\,-1/2}$ restricted to the values of Eq. (\ref{src}). This
leads to a domain of values obeying the inequality
(\ref{gsMs-range}). However, this exact spinon selection rule does
not restrict the length $\nu$ of the spin Takahashi's ideal string
excitations which have quantum overlap with
${\cal{N}}$-rotated-electron excitations
${\tilde{O}}_{\cal{N}}\vert GS\rangle$. This is because according
to the expressions given in Eq. (C25) of the companion paper
\cite{I}, the ground state occupancy of $-1/2$ spinons is finite
and given by $M^0_{s,\,-1/2}=N^0_{s,\,1}=N^0_{\downarrow}$. Thus
one can generate both $-1/2$ HL spinons and $s,\nu$
pseudoparticles belonging to branches such that $\nu >1$ by
decomposition processes of $s,1$ pseudoparticles which do not
change the net number of $-1/2$ spinons. We note that the exact
spinon selection rules just limit the value of the deviations of
such a net number and do not limit the processes which conserve
its value.

The exact selection rule (\ref{src}) is equivalent to the
inequality (\ref{gSc-range}). This latter inequality reveals that
the deviation in the number of $c$ pseudoparticle holes generated
by application onto a ground state of a rotated-electron operator
is limited by an exact selection rule. Based on the expressions
provided in the companion paper \cite{I}, one can express the
number of $c$ pseudoparticle holes and the number of $s,1$
pseudoparticle holes in terms of the values of $\eta$-spin $S_c$
and spin $S_s$ as follows,

\begin{equation}
N_c^h = 2S_c + 2\sum_{\nu =1}^{\infty}\nu\, N_{c,\,\nu} = L_c +
2\sum_{\nu =1}^{\infty}\nu\, N_{c,\,\nu} \, , \label{NhcSc}
\end{equation}
and

\begin{equation}
N_{s,\,1}^h = 2S_s + 2\sum_{\nu =1}^{\infty}(\nu -1)\, N_{s,\,\nu}
= L_s + 2\sum_{\nu =1}^{\infty}(\nu -1)\, N_{s,\,\nu} \, ,
\label{Nhs1Ss}
\end{equation}
respectively. Based on the symmetries between the charge and spin
sectors of the 1D Hubbard model, one would expect that the values
of the deviation in the number of $s,1$ pseudoparticle holes
generated by application onto a ground state of a rotated-electron
operator should also be limited. Indeed, below we find numerical
evidence that within the permitted final states which obey the
selection rules (\ref{src}) and (\ref{gSc-range}), there is a sub
class of states that describe over 94\% of the
few-rotated-electron spectral weight and whose deviation values in
the number of $s,1$ pseudoparticle holes are such that,

\begin{equation}
-\sum_{l_c,\,l_s=\pm
1}(l_c.\,l_s)\,{\cal{N}}_{l_c,\,l_s}\leq\Delta N_{s,\,1}^h\leq
{\cal{N}} \, . \label{gsNhs1-range}
\end{equation}

Note that the exact selection rule (\ref{src}) limits the number
of $-1/2$ Yang holons and $c,\nu$ pseudoparticles generated by
application onto a ground state of a ${\cal{N}}$-rotated-electron
operator ${\tilde{O}}_{\cal{N}}$. Also the states whose deviation
values obey Eq. (\ref{gsNhs1-range}) have the following
restrictions in the number of $-1/2$ HL spinons and $s,\nu$
pseudoparticles belonging to $\nu
>1$ branches,

\begin{equation}
\Delta {\breve{M}}_{s,\,-1/2} = \Delta L_{s,\,-1/2}+\sum_{\nu
=1}^{\infty}(\nu-1)\, \Delta N_{s,\,\nu} =0,1,2,...,
\sum_{l_c,l_s=\pm 1}\delta_{l_c,\,l_s}\, {\cal{N}}_{l_c,\,l_s} \,
. \label{srs}
\end{equation}
Here the quantity ${\breve{M}}_{s,\,-1/2}$ is given by,

\begin{equation}
{\breve{M}}_{s,\,-1/2} \equiv L_{s,\,-1/2}+\sum_{\nu
=1}^{\infty}(\nu-1)\, N_{s,\,\nu} = M_{s,\,-1/2} - \sum_{\nu
=1}^{\infty} N_{s,\,\nu} \, . \label{LNsn-1}
\end{equation}
Note that the value of this number vanishes for the ground state.
Equations (\ref{gsNhs1-range}) and (\ref{srs}) contain the same
information.

Let us consider the specific case of one rotated-electron removal
and addition. In this case the states obeying the deviation value
restrictions of Eqs. (\ref{gsNhs1-range}) and (\ref{srs}) are such
that $\Delta N_{s,\,1}^h =1$. Since the rotated-electron spectral
weight distributions are independent of the value of $U/t$ and as
$U/t\rightarrow\infty$ rotated electrons equal electrons, we can
find the latter distributions by evaluating the electron spectral
weight for $U/t\rightarrow\infty$. By use of the method reported
in Ref. \cite{Penc97}, we find that as the number of sites $L$
increases the one $s,1$ pseudoparticle hole excitations have for
one-electron removal and addition and $U/t\rightarrow\infty$ the
relative weights provided in the Table below. We expect that for
$L\rightarrow\infty$ the value of the relative weights for
one-electron removal and addition given in the Table for smaller
systems are as $U/t\rightarrow\infty$ above 98\% and 94\%,
respectively. The numbers provided in the Table seem to confirm
that the deviation value restrictions of Eqs. (\ref{gsNhs1-range})
and (\ref{srs}) refer to processes that lead to a substantial part
of the rotated-electron spectral weight associated with the
permitted states whose deviation values obey the exact selection
rule (\ref{src}). The $U/t\rightarrow\infty$ results of Ref.
\cite{Penc97} were derived by a scheme where the spin degrees of
freedom of the 1D Hubbard model were described by a Heisenberg
isotropic spin model. Within such a scheme the one-electron
excitations were in part simulated by a procedure involving a
change in the number of sites of the Heisenberg chain. This
procedure has similarities with the one used in Refs.
\cite{PWA,Talstra}. The deviation value restrictions of Eq.
(\ref{srs}) do not correspond to an exact selection rule but are
behind the result considered surprising in Ref. \cite{Penc97} that
for $U/t\rightarrow\infty$, more than 97\% and more than 99\% of
the spectral weight associated with creation of an electron and
annihilation of an electron, respectively, is found on the spin
branch of Faddeev and Takhtajan. This branch dispersion refers to
the part of the one-rotated-electron spectral weight associated
with the spin degrees of freedom. It is well known from the
Bethe-ansatz solution that the holes in the $c$ and $s,1$
pseudoparticle bands play a key role in the energy spectrum of the
charge and spin excitations of the 1D Hubbard model, respectively.
Note that the ground-state selection rules (\ref{src}) and
deviation value restrictions (\ref{srs}) can be expressed in terms
of the deviation values of the numbers of $c$ and $s,1$
pseudoparticle holes through the inequalities (\ref{gSc-range})
and (\ref{gsNhs1-range}), respectively. When
${{\tilde{O}}}_{{\cal{N}}}$ is the one-rotated-electron creation
or annihilation operator, according to the ground state exact
selection rule (\ref{gSc-range}) and the deviation value
restrictions of Eq. (\ref{gsNhs1-range}), the maximum number of
holes created in the $c$ pseudoparticle and $s,1$ pseudoparticle
bands respectively is one. Thus if the initial ground state has
zero spin density the single $s,1$ pseudoparticle hole indeed
corresponds to the well known spin branch dispersion of Faddeev
and Takhtajan \cite{Faddeed}.

Similar results are expected to hold for excitations generated by
other few rotated-electron operators. The numbers provided in the
table were obtained by numerical simulations and seem to confirm
that, in contrast to Eq. (\ref{src}), Eq. (\ref{srs}) does not
correspond to an exact selection rule for deviations generated by
rotated-electron operators. Note that the maximum value on the
right-hand side of Eq. (\ref{srs}) is the same as the
corresponding $\Delta M_{s,\,-1/2}$ maximum value of the general
$-1/2$ spinon number deviation inequality (\ref{Ms-range}).

\begin{table}
\begin{tabular}{rcc}
 N   &  rotated-electron removal & rotated-electron addition \\
\tableline
 6   & 0.998792 & 0.977515 \\
 8   & 0.997486 & 0.972141 \\
10   & 0.996277 & 0.968088 \\
12   & 0.995178 & 0.964847 \\
14   & 0.994176 & 0.962156 \\
16   & 0.993258 & 0.959862 \\
18   & 0.992409 & 0.957867 \\
20   & 0.991622 & 0.956105 \\
22   & 0.990886 & 0.954531 \\
24   & 0.990196 & 0.953109 \\
\end{tabular}
\caption{The relative weight of the one $s,1$ pseudoparticle hole
contributions in the $U/t\rightarrow\infty$ limit for finite size
systems of $N$ electrons. This also gives the same relative weight
in the case of excitations generated by one-rotated-electron
operators for any value of $U/t$.}
\end{table}

\subsection{DOMINANT PROCESSES FOR THE FEW-ELECTRON SPECTRAL PROPERTIES}

Here we confirm that the ground-state charge selection rule
(\ref{src}) for deviations generated by
${\cal{N}}$-rotated-electron operators also defines the dominant
holon and spinon microscopic processes that control the spectral
properties of the corresponding ${\cal{N}}$-electron operator. The
dominant processes correspond to 99.75\% - 100.00\% of the whole
electronic spectral weight and are associated with states whose
deviation values generated by application of a
${\cal{N}}$-electron operator onto a ground state obey Eq.
(\ref{src}). Within such dominant processes, there is a sub-group
of simple holon and spinon processes that correspond to about
94.00\% of the electronic spectral weight. These latter processes
are associated with states whose deviation values generated by
application of a ${\cal{N}}$-electron operator onto a ground state
obey both Eqs. (\ref{src}) and (\ref{srs}). (We recall that the
deviation value restrictions of Eq. (\ref{srs}) are not exact
selection rules for rotated-electron operators.)

We start by considering the specific case of one-electron
addition. We recall that according to the results of the companion
paper I, the number of holons $M_c =M_{c,\,-1/2} +M_{c,\,+1/2}$
and the number of spinons $M_s =M_{s,\,-1/2} +M_{s,\,+1/2}$ are
such that $M_c=[N_a-N_c]$ and $M_s=N_c$, respectively. For
simplicity, let us assume that the initial ground state has zero
spin density. In this case the spectral-weight distribution
associated with creation of a spin-up electron has the same form
as the one associated with creation of a spin-down electron. Here
we consider the former case. The spin-up electron operator is
related by Eq. (\ref{ONtil}) to the spin-up rotated electron. Thus
in the present one-electron case the selection rule (\ref{src})
and the deviation value restrictions given in Eq. (\ref{srs})
refer to creation of a spin-up rotated electron. From the use of
expressions (46), (47), (49), and (50) of the companion paper
\cite{I} for the electron and rotated-electron numbers in terms of
the holon, spinon, and $c$ pseudoparticle numbers, we find that in
the case of creation of a spin-up rotated electron the following
transitions are permitted by the selection rule (\ref{src}) and by
the equivalent inequality (\ref{gSc-range}):

(i) Transitions such that $-\Delta N_c = \Delta N^h_c=-1$, $\Delta
M_{c,\,-1/2}=0$, $\Delta M_{s,\,-1/2}=0$, $\Delta M_c =-1$, and
$\Delta M_s =1$. The minimal excitation energy for such
transitions is zero. Within these general transitions, the
transitions that also obey the restrictions (\ref{gsNhs1-range})
and (\ref{srs}) are such that $\Delta N_{s,\,1}=0$ and $\Delta
N^h_{s,\,1}=1$.

(ii) Transitions such that $-\Delta N_c = \Delta N^h_c=1$, $\Delta
M_{c,\,-1/2}=1$, $\Delta M_{s,\,-1/2}=-1$, $\Delta M_c =1$, and
$\Delta M_s =-1$. The minimal excitation energy for such
transitions is $E_u$. Within these general transitions, the
transitions that also obey the restrictions (\ref{gsNhs1-range})
and (\ref{srs}) are such that $\Delta N_{s,\,1}=-1$ and $\Delta
N^h_{s,\,1}=1$.

In the case of creation of a spin-up rotated electron the
selection rule (\ref{src}) imposes that $\Delta
M_{c,\,-1/2}=0,\,1$ and thus that $\Delta N^h_c\leq 1$. Note that
the general transitions (i) and (ii) obey such restrictions.
Within these general transitions, the transitions that also obey
the restrictions (\ref{srs}) are such that $[\Delta M_{s,\,-1/2}
-\Delta N_{s,\,1}]=0$ and thus that $\Delta N^h_{s,\,1}\leq 1$.

The simplest non-permitted transitions involve creation of three
holes in the $c$ pseudoparticle band and of three holons:

(iii) Transitions such that $-\Delta N_c = \Delta N^h_c=3$,
$\Delta M_{c,\,-1/2}=2$, $\Delta M_{s,\,-1/2}=-2$, $\Delta M_c
=3$, and $\Delta M_s =-3$. The minimal excitation energy for such
transitions is $2E_u$. These transitions are not permitted by the
rule (\ref{src}) because $\Delta M_{c,\,-1/2}=2>1$. Within these
general transitions, the transitions that obey the restrictions
(\ref{gsNhs1-range}) and (\ref{srs}) are such that $\Delta
N_{s,\,1}=-2$ and $\Delta N^h_{s,\,1}=1$.

Within the permitted transitions of type (ii), the simplest
transitions that do not obey the restrictions (\ref{gsNhs1-range})
and (\ref{srs}) involve creation of three holes in the $s,1$
pseudoparticle band:

(ii') Transitions such that $-\Delta N_c = \Delta N^h_c=1$,
$\Delta M_{c,\,-1/2}=1$, $\Delta M_{s,\,-1/2}=-1$, $\Delta M_c
=1$, and $\Delta M_s =-1$ and with $\Delta N_{s,\,1}=-2$, $\Delta
N^h_{s,\,1}=3$. The minimal excitation energy for such transitions
is $E_u$. These transitions do not obey the restrictions
(\ref{srs}) because $[\Delta M_{s,\,-1/2} -\Delta
N_{s,\,1}]=\Delta L_{s,\,-1/2}=1>0$.

In the case of half filling we often shift the ground-state
zero-energy level to the middle of the Mott-Hubbard gap. In that
case the above minimal excitation energies $0$, $E_u$, and $2E_u$
become $0$, $E_{MH}/2$, and $3E_{MH}/2$, respectively, where
$E_{MH}=E_u$ at $n=1$. Below we call {\it three-holon states} the
final states associated with the transitions (iii) because they
involve creation of three holons. These states also involve
creation of three holes in the $c$ pseudoparticle band. Moreover,
we call {\it three-$s,1$-hole states} the final states behind
transitions (ii'), which involve creation of three holes in the
$s,1$ pseudoparticle band. In contrast, the final states
associated with both the transitions (i) and (ii) are one-holon
and one $s,1$-hole states.

Let us now find out what the relative weight of the states (iii)
(or states (iii) and (ii')) is relative to the weight of the
states (i), (ii), and (ii') (or states (i) and (ii)) when all
these states are generated by application onto the ground state of
the spin-up electron creation operator. To assess the importance
of the three-holon final states (iii) and three-$s,1$-hole states
(ii'), first we turn to exact diagonalization of small chains. The
small-chain results are not expected to be a good approximation
for the evaluation of the thermodynamic-limit weight distribution
if we consider the weight associated with each specific final
state. However, here we are mostly interested in the relative
spectral-weight sum rules of the permitted states (i), (ii), and
(ii') versus the {\it forbidden} states (iii). Moreover, we also
want to know the relative weight of the states (ii'). Fortunately,
we find below that in the case of such sum rules, which involve
contributions from a whole class of states, the small-chain
results provide values for the relative weights which agree up to
99\% with the corresponding thermodynamic-limit values.

\begin{figure}[!htb]
  \centering
  \includegraphics[width=9.0truecm]{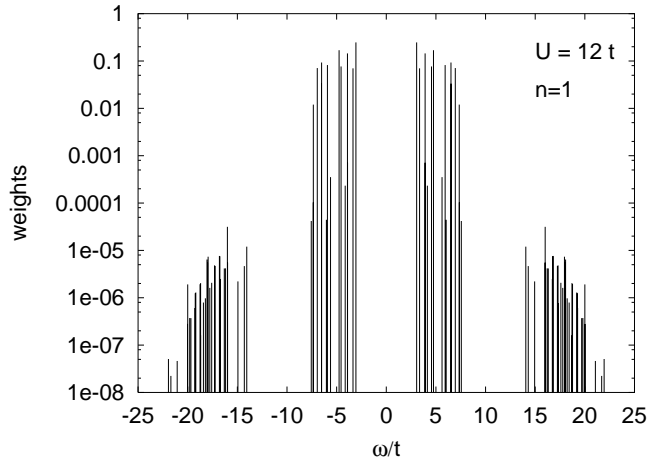}
  \caption{Electron addition ($\omega >0$) and removal ($\omega <0$)
spectrum for the half-filled six-site ring, with $U=12\,t$. Note
the logarithmic scale for the weights. Not shown are the
five-holon states, whose contributions are extremely small and
energies are out of the shown energy window.
    \label{fig:aw2}}
\end{figure}

The full electron addition and removal spectrum for six sites with
six electrons (half filling) is shown in Fig.~\ref{fig:aw2} for a
relatively large value of $U =12\,t$. As mentioned above, in the
case of half filling we define the Fermi level in the middle of
the Mott-Hubbard gap. The Hubbard bands at $\pm E_{MH}/2$ and $\pm
3E_{MH}/2$ are well separated, and the weights at $\approx \pm
3E_{MH}/2$ are orders of magnitude smaller then the contribution
of the main Hubbard bands, centered around $\pm E_{MH}/2$. The
states centered around $3E_{MH}/2$ are three-holon states of type
(iii). As a result of the half-filling particle-hole symmetry,
there is a corresponding structure for electron removal centered
around $-3E_{MH}/2$. In the large-$U/t$ limit the latter structure
is associated with creation of two real-space empty sites. For the
general $U/t$ case such a structure results from creation of two
rotated-electron empty sites. We recall that the ${\bar{M}}$
summations of the few-electron spectral functions
(\ref{ONsfB})-(\ref{ONMsfB}) refer to such a number of
rotated-electron empty sites.

\begin{figure}[!htb]
  \centering
  \includegraphics[width=9.0truecm]{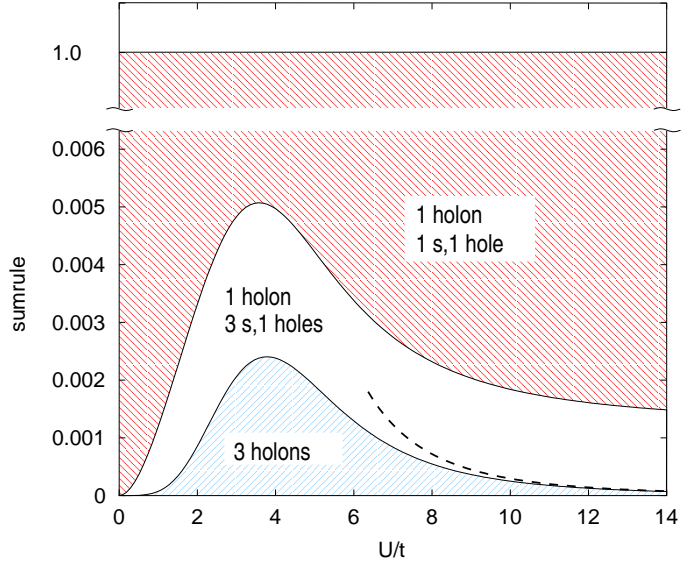}
  \caption{The contribution of different states to the sum rule. Over
  99\% of the sum rule is exhausted by the one-holon and one-$s,1$
  pseudoparticle hole excitations corresponding to the above states (i) and
  (ii). For larger systems this remains true if we consider the
  above states (i), (ii), and (ii'), which are permitted states
  for the corresponding rotated-electron problem.
    \label{fig:sumrule}}
\end{figure}

In Fig.~\ref{fig:sumrule} we have plotted the contribution of
different final states to the sum rule in the case of half
filling. For that reason, we have followed adiabatically the
weights (matrix elements) of different states as we reduced $U/t$
(which in the present case we did for this relatively small
system), and summed the weight over the particular family of
states. As we can see, the contribution of the three-holon final
states (iii) to the total sum rule is largest at intermediate
values of $U \approx 4t$, and it does not exceed 0.25\% in the
total sum rule. For large values of $U/t$ it decreases as
$(t/U)^4$, while for small values as $(U/t)^4$. The
three-$s,1$-hole contribution from the states (iv) is also less
than 0.25 \%, and for small values of $U/t$ it goes as $(U/t)^2$.
We find below that the relative spectral weight of the final
states (iii) decreases with decreasing density. For instance, at
quarter filling such a weight is 2\% of that of half filling and
vanishes in the limit of vanishing electronic density. Thus at
quarter filling and $U\approx 4t$ the final states (i), (ii), and
(ii') correspond to $\approx$ 99.99\% of the total spectral weight
and the states (iii) to $\approx$ 0.005\% of such a weight. On the
other hand, we know from the values of the above table that the
relative weight of the states (ii') increases with increasing
values of $U/t$. As we confirm below, this is not so for the
states (iii), that remain having very little spectral weight as
$L\rightarrow\infty$.

While we couldn't do finite size analysis due to lack of larger
system sizes for arbitrary value of $U$, for large $U/t$ we have
another method to calculate the contribution of the three-holon
states (iii). When $U/t$ is large, it is possible to achieve a
more precise statement through the systematic $t/U$ expansion of
the electron - rotated-electron unitary operator
\cite{Harris,Mac,Eskes}. The electron - rotated-electron canonical
transformation maps the Hamiltonian onto a rotated Hamiltonian.
Such a transformation applies to all operators, as given in Eq.
(\ref{Otil}). For instance, it also relates the elementary rotated
electron operators to the elementary electron operators when
calculating the spectral functions. So the creation operator
according to the expansions (\ref{M}) and (\ref{Mtil}) can be
decomposed into,

\begin{equation}
  c^\dagger_{j,\,\sigma} = \cdots +  c^\dagger_{j,\,\sigma;\,-E_u} +
c^\dagger_{j,\,\sigma;\,0} + c^\dagger_{j,\,\sigma;\,E_u} +
  c^\dagger_{j,\,\sigma;\,2E_u} + \cdots \, ,
\end{equation}
where the index $M$ in $c^\dagger_{j,\,\sigma;\,ME_u}$ refers to
the change of rotated-electron double occupation. Although in the
present large $U/t$ limit one has that $E_u\approx U$, we use the
general notation $ME_u$ for labelling each Hubbard band. The
operator $c^\dagger_{j,\,\sigma;\,0} =
{\tilde{c}}^\dagger_{j,\,\sigma} (1- \hat{n}_{j,\,\bar\sigma}) +
O(t/U)$ adds an electron to an unoccupied site. On the other hand,
$c^\dagger_{j,\,\sigma;\,E_u} = {\tilde{c}}^\dagger_{j,\,\sigma}
n_{j,\,\bar\sigma} + O(t/U)$ adds an electron to a site already
occupied by an electron of opposite spin, thus promoting the state
to the subspace with one more doubly occupied site. In the
$U/t\rightarrow\infty$ limit these operators change the number of
$c$ pseudoparticles and holons accordingly to the deviation values
of the above transitions (i) and (ii)-(ii'), respectively.

The operator involved in the generation of the three-holon states
(iii) is $c^\dagger_{j,\,\sigma;\,2E_u}$. Its first non-vanishing
term is proportional to $(t/U)^2$ and reads,

\begin{equation}
  \label{eq:ano}
c^\dagger_{j,\,\sigma;\,2E_u} =
  \frac{1}{U^2}
  \left(
    \left[
      \left[
        {\tilde{T}}_{E_u},{\tilde{T}}_0
      \right]
      , {\tilde{c}}^\dagger_{j,\,\sigma} {\tilde{n}}_{j,\,\bar\sigma}
    \right]
  + \frac{1}{2}
    \left[
        {\tilde{T}}_{E_u},
      \left[
        {\tilde{T}}_{E_u}
      , {\tilde{c}}^\dagger_{j,\,\sigma} (1-{\tilde{n}}_{j,\,\bar\sigma})
      \right]
    \right]
 \right),
\end{equation}
where the rotated kinetic energy operators ${\tilde{T}}_{0}$ and
${\tilde{T}}_{E_u}$ are given in Eqs. (\ref{Ttil0}) and
(\ref{Ttil+1}), respectively. After some algebra we find,

\begin{eqnarray}
c^\dagger_{j,\,\sigma;\,2E_u} &=& \frac{t^2}{U^2} \Bigl[
     {\tilde{c}}^{\dagger}_{j+1,\,\sigma}   {\tilde{c}}^{\dagger}_{j-1,\,\bar\sigma}
      {\tilde{c}}^{\phantom{\dagger}}_{j,\, \bar\sigma}  \hat n_{j-1,\, \sigma}
      (1 \!-\! \hat n_{j,\, \sigma})  \hat n_{j+1,\, \bar\sigma}
\nonumber\\
 && -  {\tilde{c}}^{\dagger}_{j+1,\, \sigma}   {\tilde{c}}^{\dagger}_{j,\, \bar\sigma}
  {\tilde{c}}^{\phantom{\dagger}}_{j-1,\, \bar\sigma}  (1 \!-\! \hat n_{j-1,\, \sigma})
  \hat n_{j,\, \sigma}  \hat n_{j+1,\, \bar\sigma}
\nonumber\\
 && -  {\tilde{c}}^{\dagger}_{j,\, \sigma}   {\tilde{c}}^{\dagger}_{j+1,\, \bar\sigma}
  {\tilde{c}}^{\phantom{\dagger}}_{j-1,\, \bar\sigma}  (1 \!-\! \hat n_{j-1,\, \sigma})
  \hat n_{j,\, \bar\sigma}  \hat n_{j+1,\, \sigma}
\nonumber\\
 && -  {\tilde{c}}^{\dagger}_{j,\, \sigma}   {\tilde{c}}^{\dagger}_{j-1,\, \bar\sigma}
  {\tilde{c}}^{\phantom{\dagger}}_{j+1,\, \bar\sigma}  \hat n_{j-1,\, \sigma}  \hat n_{j,\, \bar\sigma}
  (1 \!-\! \hat n_{j+1,\, \sigma})
\nonumber\\
 && +  {\tilde{c}}^{\dagger}_{j-1,\, \sigma}   {\tilde{c}}^{\dagger}_{j+1,\, \bar\sigma}
  {\tilde{c}}^{\phantom{\dagger}}_{j,\, \bar\sigma}  \hat n_{j-1,\, \bar\sigma}
  (1 \!-\! \hat n_{j,\, \sigma})  \hat n_{j+1,\, \sigma}
\nonumber\\
 && -  {\tilde{c}}^{\dagger}_{j-1,\, \sigma}   {\tilde{c}}^{\dagger}_{j,\, \bar\sigma}
  {\tilde{c}}^{\phantom{\dagger}}_{j+1,\, \bar\sigma}  \hat n_{j-1,\, \bar\sigma}
  \hat n_{j,\, \sigma}  (1 \!-\! \hat n_{j+1,\, \sigma})
  \Bigr] \, .
\end{eqnarray}
The operator only acts when all the $j-1$, $j$, and $j+1$ sites
are singly occupied. The action of
$c^\dagger_{j,\,\uparrow;\,2E_u}$ on sites $j-1,j,j+1$, apart from
the $t^2/U^2$ coefficient, gives

\begin{eqnarray}
 | \uparrow \downarrow \downarrow \rangle &\rightarrow&
 | dde \rangle +  | ded \rangle
\nonumber\\
 | \downarrow \uparrow \downarrow \rangle &\rightarrow&
 - | dde \rangle +  | edd \rangle
\nonumber\\
 | \downarrow \downarrow \uparrow \rangle &\rightarrow&
 -| ded \rangle -  | edd \rangle
\nonumber
\end{eqnarray}
while on any other configuration it will give zero. Here the
indexes $d$ and $e$ stand for doubly-occupied sites and empty
sites, respectively, and the $\uparrow$ and $\downarrow$ symbols
refer to spin-up and spin-down singly-occupied sites,
respectively. We can see that the operator
$c^\dagger_{j,\,\uparrow;\,2E_u}$ removes three spins with total
spin S=1/2 (for the S=3/2 combination $| \uparrow \downarrow
\downarrow \rangle +
 | \downarrow \uparrow \downarrow \rangle +
 | \downarrow \downarrow \uparrow \rangle \rightarrow 0$
 ).

Next, we calculate $\int \mathcal{A}^{\rm 2UHB}(\omega)\, d
\omega$, the total weight in the {\it band} centered around
$3E_u/2\approx 3U/2 $ and of energy width $12\,t$ (each of the $e$
and $d$ sites contributes $4t$ to the bandwidth). It is given by
the expectation value $\sum_\sigma \langle \psi_0\vert\,
c^{\phantom{\dagger}}_{j,\,\sigma;\,-2E_u}
c^\dagger_{j,\,\sigma;\,2E_u}\vert\, \psi_0 \rangle$, where the $|
\psi_0 \rangle$ is the spin-charge factorized wave function
\cite{Ogata}. First we find that,

\begin{equation}
  \label{eq:aasn}
\sum_\sigma c^{\phantom{\dagger}}_{j,\,\sigma;\,-2E_u}
c^\dagger_{j,\,\sigma;\,2E_u} =\frac{t^4}{U^4}
  \left( \frac{3}{2} -2 {\bf S}_{j-1} {\bf S}_{j}
  -2 {\bf S}_{j} {\bf S}_{j+1}-2 {\bf S}_{j-1} {\bf S}_{j+1}   \right)
\hat {\cal{P}}_{1,j-1}\, \hat {\cal{P}}_{1,j}\, \hat
{\cal{P}}_{1,i+j} \, ,
\end{equation}
where the projector $\hat {\cal{P}}_{1,j} = \hat n_{j,\,\uparrow}
+ \hat n_{j,\,\downarrow} - 2 \hat n_{j,\,\uparrow} \hat
n_{j,\,\downarrow}$ ensures the single occupation of site $j$.

Replacing the factorized wave function, the partial sum rule
reads,

\begin{equation}
 \int \mathcal{A}^{\rm 2UHB}(\omega)\, d \omega   =
  \left\langle \frac{3}{2} -2 {\bf S}_0 {\bf S}_1
  -2 {\bf S}_1 {\bf S}_2-2 {\bf S}_0 {\bf S}_2   \right\rangle_{\rm Heis}
\langle \hat n_{0} \hat n_1 \hat n_2 \rangle_{\rm sf}
\frac{t^4}{U^4} \, .
\end{equation}

Let us denote by $f^{\dagger}_{j}$ and $f_{j}$ the creation and
annihilation operators, respectively, of a spin-less fermion at
site $j$. The expectation value to find three neighboring spinless
fermions is

\begin{eqnarray}
\langle \hat n_{0} \hat n_{1} \hat n_{2} \rangle_{\rm sf} &=&
\left|
\begin{array}[c]{ccc}
\langle f^{\dagger}_{0} f^{\phantom{\dagger}}_{0}  \rangle &
\langle f^{\dagger}_{0} f^{\phantom{\dagger}}_{1}  \rangle &
\langle f^{\dagger}_{0} f^{\phantom{\dagger}}_{2}  \rangle \\
\langle f^{\dagger}_{1} f^{\phantom{\dagger}}_{0}  \rangle &
\langle f^{\dagger}_{1} f^{\phantom{\dagger}}_{1}  \rangle &
\langle f^{\dagger}_{1} f^{\phantom{\dagger}}_{2}  \rangle \\
\langle f^{\dagger}_{2} f^{\phantom{\dagger}}_{0}  \rangle &
\langle f^{\dagger}_{2} f^{\phantom{\dagger}}_{1}  \rangle &
\langle f^{\dagger}_{2} f^{\phantom{\dagger}}_{2}  \rangle
\end{array}
\right| \nonumber\\
&=& n^3 - \frac{2 n \sin^2(\pi n)}{\pi^2} +
  \frac{\sin^2(\pi n) \sin(2 \pi n)}{\pi^3} -
  \frac{n \sin^2(2 \pi n)}{4 \pi^2}
\end{eqnarray}
with the limiting behavior,

\begin{equation}
  \label{eq:nnnser}
  \langle \hat n_{0} \hat n_{1} \hat n_{2} \rangle_{\rm sf} =
\left\{
\begin{array}[c]{ll}
\displaystyle{\frac{4\pi^6}{135}}n^9 &\mbox{if $n \ll 1$ ;} \\
& \\
1-3(1-n) &\mbox{if $1-n \ll 1$ .}
\end{array}
\right. \, .
\end{equation}
Note that the spectral weigh of these states decreases rapidly
away from half filling, {\it i.e.} at quarter filling ($n=1/2$) it
is about 2\% of that at half filling.

The expectation values in the thermodynamic limit of the
Heisenberg model is \cite{takahashi77},

\begin{eqnarray}
   \langle {\bf S}_0 {\bf S}_1 \rangle_{\rm Heis} &=& \frac{1}{4}-\ln 2
    \approx -0.443147\\
   \langle {\bf S}_0 {\bf S}_2 \rangle_{\rm Heis} &=& \frac{1}{4}-4 \ln 2
   +\frac{9}{4}\zeta(3) \approx 0.182039
\end{eqnarray}
so that
\begin{equation}
  \label{eq:exps1o2}
  \left\langle \frac{3}{2} -2 {\bf S}_0 {\bf S}_1
  -2 {\bf S}_1 {\bf S}_2-2 {\bf S}_0 {\bf S}_2   \right\rangle_{\rm Heis}
  = 12 \ln 2 - \frac{9}{2} \zeta(3) \approx 2.91 \, .
\end{equation}
The sum rule in the second upper Hubbard band defined in Sec. III
is then,

\begin{equation}
 \int \mathcal{A}^{\rm 2UHB}(\omega)\, d \omega   \approx 2.91\,
  \langle \hat n_{0} \hat n_{1} \hat n_{2} \rangle_{\rm sf}\,
\frac{t^4}{U^4} \, .
\end{equation}

For the six-site finite-size cluster, the expectation value in
Eq.~(\ref{eq:exps1o2}) is $(169+17 \sqrt{13})/78 \approx 2.95$,
which is about 1\% off from the thermodynamic-limit value. The
asymptotic $2.95\,t^4/U^4$ is shown in Fig.~\ref{fig:sumrule} as a
dashed line.

Both our numerical results for general values of $U/t$ and a small
system and our analytical large $U/t$ results for the
thermodynamic limit confirm that the ground-state selection rule
(\ref{src}) which limits the values of the deviations generated by
rotated-electron operators defines the dominant holon and spinon
microscopic processes that control the spectral-weight
distribution of the corresponding electronic operators. While for
$U<<t$ and $U>>t$ such a selection rule is exact also for these
electronic operators, we find that the relative spectral weight of
the {\it permitted} above states (i), (ii), and (ii') is minimum
for $U\approx 4t$. This {\it minimum} value decreases with
decreasing density. For half filling it is given by $\approx$
99.75\%, whereas for quarter filling it reads $\approx$ 99.99\%
and in the limit of vanishing density it becomes $\approx$
100.00\%. The extremely small amount of missing spectral weight
corresponds mainly to the {\it forbidden} three-holon states
(iii). Higher order five-holon/five-$c$-hole states lead to nearly
vanishing spectral weight.

While the selection rule (\ref{src}) is exact for rotated-electron
operators, the states obeying the deviation value restrictions of
Eq. (\ref{srs}) correspond to rotated-electron processes that
generate most spectral weight of the permitted transitions.
Although the relative spectral weight of the three-$s,1$-hole
states (ii') increases with increasing the system length, we
expect that for one-electron addition its maximum value is about
6\% as $L\rightarrow\infty$ for half filling. However, higher
order five-$s,1$-hole states lead to nearly vanishing electronic
spectral weight. Thus over 99\% of the one-electron addition
spectral weight corresponds to generation of one holon and one and
three $s,1$ pseudoparticle holes.

A similar analysis can be performed for other few-electron
operators. The deviation restrictions of Eq. (\ref{src}) also
amount to more than 99\% of the spectral weight generated by
application onto the ground state of ${\cal{N}}$-electron
operators such that ${\cal{N}}>1$. Importantly, for all
few-electron operators these deviation restrictions become exact
both for $U<<t$ and $U>>t$ and refer to the dominant holon and
spinon microscopic processes for $U\approx 4t$. The spectral
weight associated with excited states whose deviation values are
outside the ranges defined by Eq. (\ref{src}) is for electronic
operators in general maximum for half filling, yet it remains
small for such an electronic density. As an example we consider
the two-electron frequency dependent optical conductivity, which
is directly related to the dynamical structure factor. A
preliminary finite-energy study of such a ${\cal{N}}=2$ electron
problem was presented for the case of the 1D Hubbard model in Ref.
\cite{optical}. The ground-state selection rule of Eq. (\ref{src})
implies that in that case the corresponding two-rotated-electron
problem the permitted $-1/2$ holon number deviations are such that
$\Delta M_{c,-1/2}=0,\,1$ and thus the permitted holon number
deviations are $\Delta M_c =0,\,2$, respectively. The
finite-energy absorption considered in that paper results from
transitions associated with holon number deviations such that
($\Delta M_{c,-1/2}=0;\,\Delta M_c=0$) and ($\Delta
M_{c,-1/2}=1;\,\Delta M_c=2$). Thus contributions from excited
states with holon number deviations such that ($\Delta
M_{c,-1/2}=2;\,\Delta M_c=4$) are expected to correspond to less
than 1\% of the two-electron optical conductivity spectral weight.

As for the one-electron problem, it is expected that such an
extremely small amount of spectral weight is maximum yet very
small at half filling. Such a prediction is confirmed by the
results of Ref. \cite{Controzzi}. There such a finite-energy
absorption was investigated for the sine-Gordon model. The
absorptions represented for coupling constant $\beta^2=0.9$ by a
solid line and a dashed line in Fig. 1 of the same reference
correspond to one-soliton/one-antisoliton and
two-solition/two-antisoliton contributions, respectively. The
latter absorption is very small, being multiplied by $100$ in the
figure. In the limit of $\beta^2\rightarrow 1$ the sine-Gordon
model acquires a $\eta$-spin $SU(2)$ symmetry and describes the 1D
Hubbard model at half filling and small values of $U/t$. Moreover,
in this limit the solitons and antisolitons become $+1/2$ holons
and $-1/2$ holons, respectively. Therefore, the results of that
reference are expected to be very similar to those of the
half-filling Hubbard model. As $\beta^2\rightarrow 1$ the above
one-soliton/one-antisoliton and two-solition/two-antisoliton
transitions correspond to ($\Delta M_{c,+1/2}=1;\,\Delta M_c=2$)
and ($\Delta M_{c,-1/2}=2;\,\Delta M_c=4$) deviations,
respectively. The transitions associated with the latter
deviations are forbidden for the corresponding
two-rotated-electron operator. Thus in the case of the
two-electron operator the contributions from these transitions are
expected to amount for less than 1\% of the spectral weight. Such
a prediction is confirmed by the smallness of the corresponding
absorption, which is represented in Fig. 1 of Ref.
\cite{Controzzi}. Similar results hold for other two-electron
operators.

We emphasize that equivalent results also hold for one-electron
removal. It is generally accepted that one-electron removal
corresponds to final states with a single extra hole in the $c$
pseudoparticle band. However, we find that this general
expectation is only true for one-rotated-electron removal. In the
case of one-electron removal less than 1\% of the spectral weight
goes again to final states with three extra $c$ pseudoparticle
holes. In the particular case of half filling such final states
lead to the structure of Fig.~\ref{fig:aw2} which is centered
around $-3E_{MH}/2$.

\section{DISCUSSION AND CONCLUDING REMARKS}

In this section we summarize and discuss the results obtained in
this paper. Moreover, we discuss the application of the concepts
and non-perturbative many-electron tools introduced here and in
the companion papers \cite{I,III} to the accomplishment of a
program for the study of few-electron spectral functions at finite
values of excitation energy. Such an application program is
fulfilled in Refs. \cite{IIIb,V}.

\subsection{SUMMARY OF THE RESULTS}

The holon, spinon, and $c$ pseudoparticle description introduced
in the first paper \cite{I} of this series and further studied in
this second and in the third \cite{III} papers, can be used in the
evaluation of matrix elements between the ground state and excited
states and in the related problem of the derivation of the line
shape for finite-energy few-electron spectral functions
\cite{IIIb,V}. The results of the present paper about the dominant
microscopic physical processes that amount for more than 99\% of
the spectral weight generated by application onto the ground state
of few-electron operators is useful for the success of such a
program. Our theory also describes the higher-order holon and
spinon processes associated with the remaining less than 1\% of
electronic spectral weight. Furthermore, in this paper we used the
relation of rotated electrons to the quantum objects whose
occupancy configurations describe the energy eigenstates, in the
generalization of the concepts of lower Hubbard band and upper
Hubbard bands for all values of the on-site repulsion. While the
lower Hubbard band refers to zero rotated-electron double
occupation final states, the $Mth$ upper Hubbard band corresponds
to the spectral weight associated with the excited states of
rotated-electron double occupation $M$. For large values of the
ratio $U/t$ this definition of lower Hubbard band and upper
Hubbard bands coincides with the usual one.

According to the inequalities given in Eq. (\ref{DMcsrange}), the
maximum number of holons and spinons generated or annihilated by
application of $\cal{N}$-rotated-electron operators onto any
eigenstate of the spin $\sigma$ electron number operator is given
by $\cal{N}$. In the particular case of excitations whose initial
state is the ground state, the ranges provided in Eq. (\ref{src})
refer to an exact selection rule for the deviation values
generated by application onto the ground state of such
${\cal{N}}$-rotated-electron operators. The occurrence of this
selection rule is a direct result of the relation of rotated
electrons to the quantum objects whose occupancy configurations
describe all energy eigenstates of the model. In this paper we
found that the excited states associated with these permitted
deviation values correspond to over 99\% of the spectral weight
generated by application onto the same state of the corresponding
${\cal{N}}$-electron operators. This means that over 99\% of the
spectral weight generated by application onto the ground state of
a ${\cal{N}}$-electron operator corresponds to application onto
such a state of the first operator term of the second expression
of Eq. (\ref{ONtil}). Such an operator term is nothing but the
${\cal{N}}$-rotated-electron operator associated with the
${\cal{N}}$-electron operator under consideration. This
interesting result reveals that application onto the ground sate
of an operator of the form given in Eq. (\ref{DONtil}) leads to
very little spectral weight. The value $\cal{N}$ also determines
the absolute maximum value $\nu={\cal{N}}$ of the quantum number
$\nu$ of the $c ,\nu$ pseudoparticle branches which can have
finite occupancy in final states generated by
$\cal{N}$-rotated-electron operators. The number $\nu$ is also the
length of the ideal charge Takahashi's string excitations
associated with the $c,\nu$ pseudoparticle branch. Therefore, the
$\cal{N}$-rotated-electron excitations only couple to energy
eigenstates described by $c$ Takahashi's ideal string charge
excitations of length $\nu$ such that $\nu \leq {\cal{N}}$. For
the corresponding $\cal{N}$-electron excitations, the excited
states described by $c$ Takahashi's ideal string charge
excitations of length $\nu$ such that $\nu > {\cal{N}}$ amount to
less than 1\% of the electronic spectral weight. It follows that
the dominant processes generated by application onto the ground
state of a ${\cal{N}}$-electron operator originate a number of
$2\nu$-holon composite $c,\nu$ pseudoparticles and $-1/2$ Yang
holons and a number of $c$ pseudoparticle holes whose values are
within the ranges provided in Eqs. (\ref{src}) and
(\ref{gSc-range}), respectively. We recall that for electronic
densities $n\leq 1/a$ there are no $2\nu$-holon composite $c,\nu$
pseudoparticles and no $-1/2$ Yang holons in the ground state.
Higher order processes originating a number of these quantum
objects outside the range defined by Eqs. (\ref{src}) and
(\ref{gSc-range}) correspond to less than 1\% of the electronic
spectral weight. These processes are generated by application onto
the ground state of the operator given in Eq. (\ref{DONtil}).

Moreover, although the deviation value ranges provided in Eqs.
(\ref{gsNhs1-range}) and (\ref{srs}) are not associated with an
exact selection rule for rotated-electron operators, within the
dominant processes that include all values for these deviations,
they refer to quite important processes for the electronic
spectral weight. These equations refer to ranges for the numbers
of generated or annihilated $s,1$ pseudoparticle holes and for the
numbers of generated $2\nu$-spinon composite $s,\nu$
pseudoparticles such that $\nu>1$ and $-1/2$ HL spinons,
respectively. We note that for spin densities $m<n$ there are no
$2\nu$-spinon composite $s,\nu$ pseudoparticles such that $\nu>1$
and no $-1/2$ HL spinons in the ground state. For instance, in the
case of one-electron addition considered in the previous section,
Eq. (\ref{gsNhs1-range}) tells us that these important processes
involve generation of one $s,1$ pseudoparticle hole. We found that
these processes lead to about 94\% of the electronic spectral
weight, whereas excited states involving the creation of three
$s,1$ pseudoparticle holes amount for more than 5\% of such a
spectral weight. We thus conclude that microscopic processes
involving generation of more than three $s,1$ pseudoparticle holes
lead to almost no electronic spectral weight. Although in the case
of the ideal spin Takahashi's string excitations associated with
the $s,\nu$ pseudoparticle branches the ranges of Eq. (\ref{srs})
are not exact, they correspond to a sub-class of excited states
that amount to an important part of the electronic spectral
weight.

The occurrence of dominant holon and spinon microscopic processes
for the electronic spectral weight and how fast the contribution
of higher order processes vanishes is related to a property found
in Ref. \cite{V} that plays a central role in the evaluation of
the line shape for finite-energy few-electron spectral functions.
While the dominant holon and spinon microscopic processes
associated with Eqs. (\ref{src}) and (\ref{gSc-range}) amount for
more than 99\% of the few-electron spectral weight, the remaining
less than 1\% of spectral weight is mostly associated with final
states involving generation of a few more $c,\nu$ pseudoparticles,
$-1/2$ Yang holons, and $c$ pseudoparticle holes relative to the
maximum value of the ranges provided in these equations. Also the
contribution to the electronic spectral weight from excited states
involving the generation of $s,\nu$ pseudoparticles such that
$\nu>1$, $-1/2$ HL spinons, and $s,1$ pseudoparticle holes
decreases rapidly as the number of these quantum objects
increases. It follows that excited states involving creation of a
large number of the above-mentioned quantum objects lead to
vanishing spectral weight. In particular, in the present
thermodynamic limit the spectral weight of excited states whose
generation involves an infinite number of quantum-object
elementary processes vanishes exactly. Thus only processes
involving generation of a finite number of quantum objects
contribute to the few-electron spectral weight. This defines the
Hilbert subspace of few-electron excitations. The pseudofermion
description introduced in Ref. \cite{IIIb} refers to such a
Hilbert subspace. The method for evaluation of matrix elements
between the ground state and excited states introduced in Refs.
\cite{IIIb,V} takes into account all possible final states
generated by a final number of quantum-object processes, as
discussed below.

\subsection{THE EVALUATION OF FINITE-ENERGY FEW-ELECTRON SPECTRAL FUNCTIONS}

The holon, spinon, and pseudoparticle description introduced in
the companion paper \cite{I} and further studied here and in the
companion paper \cite{III} is applied in Refs. \cite{IIIb,V} to
the study of the few-electron spectral functions. The studies of
these references provide the few-electron spectral function line
shapes for all values of excitation energy. The above description
is a necessary condition for the successful fulfilment of the
program for evaluation of few-electron spectral functions at
finite values of the excitation energy. Such a program is
fulfilled in Refs. \cite{IIIb,V} by application of the new
concepts and paradigms introduced in this paper and in its
companion papers \cite{I,III} as follows: \vspace{0.25cm}

i)- The first step of the evaluation of a few-electron spectral
function of the general form (\ref{ONsf})-(\ref{ONMsf}) or
(\ref{ONsfB})-(\ref{ONMsfB}) is the expression of the
corresponding few-electron operator as the expansion in terms of
rotated-electron operators given by the second expression of Ref.
(\ref{ONtil}); \vspace{0.25cm}

ii) - Next, one uses the results of the companion paper \cite{III}
and express the rotated-electron operators in terms of local
pseudoparticles operators, Yang-holon operators, and HL-spinon
operators. Note that after use of the relation (\ref{ONtil}), the
problem of the evaluation of the matrix elements of the
few-electron spectral functions (\ref{ONMsf}) and (\ref{ONMsfB})
is equivalent to the computation of matrix elements involving the
elementary creation and annihilation operators of these quantum
objects ; \vspace{0.25cm}

iii) - However, rather than in terms of the interacting
pseudoparticles, it is more appropriate to use in the evaluation
of such matrix elements the related non-interacting operational
pseudofermion description introduced in Ref. \cite{IIIb}.
Fortunately, the few-electron spectral functions can be expressed
as a convolution of the pseudofermion spectral functions
corresponding to each pseudofermion branch with finite occupancy
in the excited states $\vert M,j\rangle$ and $\vert
{\bar{M}},j\rangle$ of the spectral-function expressions
(\ref{ONsf})-(\ref{ONMsf}) and (\ref{ONsfB})-(\ref{ONMsfB}),
respectively; \vspace{0.25cm}

iv) - Finally, the evaluation of the spectral function for each
pseudofermion branch with finite occupancy for these excited
states uses the method applied to $U/t\rightarrow\infty$ spin-less
fermion operators in Ref. \cite{Penc97}. The energy spectra of the
above excited states appearing in the spectral-function
expressions (\ref{ONMsf}) and (\ref{ONMsfB}) is obtained by direct
use of the Bethe-ansatz solution and $\eta$-spin and spin
symmetries. The state summations of these spectral-function
expressions is simplified by the fact that the absolute value of
the matrix elements vanishes rapidly as the number of
pseudofermion processes involved in the generation from the ground
state of the corresponding excited states increases. Such a rapid
vanishing is an important property for the evaluation of the
finite-energy few-electron spectral functions. It is directly
related to the occurrence of dominant quantum-object microscopic
processes. \vspace{0.25cm}

The use of the relation (\ref{ONtil}) in the evaluation of
finite-energy few-electron spectral functions \cite{IIIb,V} is a
consequence of the breakthrough of the companion paper \cite{I}
concerning the following two issues: First, the clarification of
the relation of rotated electrons to the Bethe-ansatz quantum
numbers that label the energy eigenstates; Second, the association
of these quantum numbers with objects whose occupancy
configurations describe the energy eigenstates. The occurrence of
dominant processes found in this paper is also important for the
studies of Refs. \cite{IIIb,V}. In addition to contributing to the
understanding of the non-perturbative microscopic physical
mechanisms that control the few-electron spectral properties, our
results are related to the rapid vanishing of the matrix elements
as the number of processes that generate the corresponding excited
states increases, as discussed above.

\subsection{FINAL DISCUSSION AND CONCLUDING REMARKS}

The complexity of the finite-energy one-electron and two-electron
physics of the non-perturbative 1D Hubbard model explains why
except for $U/t\rightarrow\infty$ investigations
\cite{Penc95,Penc96,Penc97}, there have not been many previous
studies about this interesting problem. In this paper we found
that excited states generated from the ground state by microscopic
processes leading to deviation values in the ranges defined by Eq.
(\ref{src}) lead to more than 99\% of the few-electron spectral
weight. Moreover, a careful comparison of the line shape of the
one-electron spectral weight for the trivial $U/t\rightarrow 0$
limit with the one obtained in Refs. \cite{Penc95,Penc96,Penc97}
for the limit $U/t\rightarrow\infty$, seems to indicate that most
spectral weight is located in the vicinity of {\it pseudoparticle
branch lines}. Such lines are generated by a set of final states
where one pseudoparticle or pseudoparticle hole is created for all
its available band-momentum values $q$ and all remaining
pseudoparticles or pseudoparticle holes are created at their {\it
Fermi points}. In addition to these elementary processes, the line
shape in the vicinity of these branch lines is for electronic and
spin densities within the ranges $0<n<1/a$ and $0<m<n$,
respectively, generated by pseudoparticle - pseudoparticle hole
processes in the $c$ and $s,1$ bands. The one-electron
spectral-weight distribution obtained in Ref. \cite{Senechal} by
numerical simulations refers to intermediate values of the ratio
$U/t$ such that $U/t=4$ and displays charge and spin branch lines,
in agreement with our prediction. Furthermore, this expectation is
confirmed for both one-electron and few-electron spectral
functions by the exact results of Ref. \cite{V}.

Consideration of all possible branch lines constructed in this way
for the dominant excited states whose deviation values obey the
ranges defined by Eqs. (\ref{src}) and (\ref{srs}) for a given
few-electron operator is expected to describe the main spectral
lines observed in real quasi-one-dimensional materials. This
expectation is confirmed in the case of the organic conductor
TTF-TCNQ by the preliminary application of our results presented
in Ref. \cite{spectral0}. The separated one-electron charge and
spin spectral lines observed in TTF-TCNQ by photoemission studies
agree for the whole energy band width with the theoretical branch
lines constructed by the above described procedure. These
preliminary results confirm the interest for the study of the
unusual finite-energy spectral properties of these materials of
the concepts introduced in this paper.

Elsewhere the powerful method introduced in Refs. \cite{IIIb,V} is
used in a more detailed description of the line shape observed in
TTF-TCNQ \cite{spectral}. While the preliminary results of Ref.
\cite{spectral0} provide the form of the charge and spin branch
lines, the detailed dependence on the excitation energy $\omega$
of the line shape in the vicinity of these lines is presented in
Ref. \cite{spectral}. Fortunately, this more detailed study also
agrees with the line shape observed in the real experiment.
Elsewhere the method constructed in Refs. \cite{IIIb,V} by means
of the concepts and non-perturbative many-electron theoretical
tools introduced here and in the companion papers \cite{I,III} is
applied to the study of the two-electron dynamical structure
factor for all values of $U/t$ \cite{dynamical}. Moreover, a study
of the phase diagram of a system of weakly coupled Hubbard chains
which combines our 1D results with a Renormalization Group scheme
will be also implemented \cite{phases}. The results of these
studies describe many of the anomalous spectral properties
observed in real low-dimensional materials. In particular, they
lead to spectral lines similar to the ones observed in
finite-energy/frequency experimental investigations of the
one-electron spectral weight distribution or organic metals
\cite{Ralph,spectral0} and two-electron dynamical structure factor
of Mott-Hubbard insulators \cite{Hasan}. Our theoretical
predictions also describe and successfully explain the microscopic
mechanisms behind the phase phase diagram observed in
quasi-one-dimensional materials \cite{Bourbonnais}.

Finally, the further application of the results obtained in this
paper and in its companion paper \cite{I} to the explicit
evaluation of finite-energy few-electron spectral functions
requires the introduction of the concepts of {\it local
pseudoparticle} and {\it effective pseudoparticle lattice}. The
introduction of these concepts is the main goal of the third and
last paper of this series, Ref. \cite{III}.

\begin{acknowledgments}
We thank Jim W. Allen, Ant\^onio Castro Neto, Ralph Claessen,
Francisco (Paco) Guinea, Katrina Hibberd, Peter Horsch, Jo\~ao
Lopes dos Santos, Lu\'{\i}s Miguel Martelo, and Pedro Sacramento
for stimulating discussions. We also thank the support of EU money
Center of Excellence ICA1-CT-2000-70029 and one of us (K.P.)
thanks the
support of OTKA grants D032689, T038162, and the Bolyai
fellowship.
\end{acknowledgments}



\begin{references}
\bibitem[1]{I}
        J. M. P. Carmelo, K. Penc, and J. M. Rom\'an, submitted for publication in Phys.
        Rev. B (2002) [cond-mat/0302044].
\bibitem[2]{III}
        J. M. P. Carmelo, submitted for publication
        in Phys. Rev. B (2002) [cond-mat/0305347].
\bibitem[3]{Lieb}
        Elliott H. Lieb and F. Y. Wu, Phys. Rev. Lett. {\bf 20},
        1445 (1968).
\bibitem[4]{Takahashi}
        M. Takahashi, Prog. Theor. Phys. {\bf 47}, 69 (1972).
\bibitem[5]{HL}
        O. J. Heilmann and E. H. Lieb, Ann. N. Y. Acad. Sci. {\bf 172},
        583 (1971); E. H. Lieb, Phys. Rev. Lett. {\bf 62}, 1201 (1989).
\bibitem[6]{Yang89}
        C. N. Yang, Phys. Rev. Lett. {\bf 63}, 2144 (1989).
\bibitem[7]{Martins98}
        P. B. Ramos and M. J. Martins , J. Phys. A {\bf 30}, L195 (1997);
        M. J. Martins and P.B. Ramos, Nucl. Phys. B {\bf 522}, 413 (1998).
\bibitem[8]{II}
        J. M. P. Carmelo and P. D. Sacramento, Phys. Rev. B {\bf 67} (2003) [cond-mat/0302042].
\bibitem[9]{Rasetti}
        M. Rasetti, {\it The Hubbard Model, Recent Results},
        Word Scientific, Singapore (1991); A. Montorsi,
        {\it The Hubbard Model}, Word Scientific, Singapore
        (1992); {\it The Hubbard Model -- Its Physics and Mathematical
        Physics}, edited by Dionys Baeriswyl, David K. Campbell,
        Jos\'e M. P. Carmelo, Francisco Guinea, and Enrique
        Louis (NATO ASI Series B, Vol. 343, Plenum Press,
        New York, 1995).
\bibitem[10]{Harris}
        A. Brooks Harris and Robert V. Lange, Phys. Rev. {\bf 157},
        295 (1967).
\bibitem[11]{Mac}
        A. H. MacDonald, S. M. Girvin, and D. Yoshioka,
        Phys. Rev. B {\bf 37}, 9753 (1988).
\bibitem[12]{Eskes}
        Henk Eskes and Andrzej M. Ole\'s, Phys. Rev. Lett. {\bf 73}, 1279
        (1994); Henk Eskes, Andrzej M. Ole\'s, Marcel B. J. Meinders, and
        Walter Stephan, Phys. Rev. B {\bf 50}, 17 980 (1994).
\bibitem[13]{Geb}
        F. Gebhard, K. Bott, M. Scheidler, P. Thomas, and S. W. Koch,
        Phylo. Magaz. B {\bf 75}, 13 (1997).
\bibitem[14]{Beni}
        G. Beni, T. Holstein, and P. Pincus, Phys. Rev. B {\bf 8},
        312 (1973).
\bibitem[15]{Klein}
        Douglas J. Klein, Phys. Rev. B {\bf 8}, 3452 (1973).
\bibitem[16]{Carmelo88}
        Jos\'e Carmelo and Dionys Baeriswyl,
        Phys. Rev. B {\bf 37}, 7541 (1988).
\bibitem[17]{Ogata}
        F. Woynarovich, J. Phys. C {\bf 15}, 97 (1982);
        Masao Ogata and Hiroyuki Shiba, Phys. Rev. B {\bf 41},
        2326 (1990).
\bibitem[18]{Parola}
        S. Sorella and A. Parola, J. Phys. Condens. Matter {\bf 4},
        3589 (1992).
\bibitem[19]{Ricardo}
        R. G. Dias and J. M. Lopes dos Santos, J. de Physique I {\bf 2},
        1889 (1992).
\bibitem[20]{Penc95}
        Karlo Penc, Fr\'ed\'eric Mila, and Hiroyuki Shiba,
        Phys. Rev. Lett. {\bf 75}, 894 (1995).
\bibitem[21]{Penc96}
        Karlo Penc, Karen Hallberg, Fr\'ed\'eric Mila, and Hiroyuki Shiba,
        Phys. Rev. Lett. {\bf 77}, 1390 (1996).
\bibitem[22]{Penc97}
        Karlo Penc, Karen Hallberg, Fr\'ed\'eric Mila, and Hiroyuki Shiba,
        Phys. Rev. B {\bf 55}, 15 475 (1997).
\bibitem[23]{Pines}
        D. Pines and P. Nozi\`eres, in {\em The Theory of
        Quantum Liquids},
        (Addison-Wesley, Redwood City, 1966 and 1989), Vol. I.
\bibitem[24]{Baym}
        Gordon Baym and Christopher J. Pethick, in
        {\em Landau Fermi-Liquid Theory Concepts and Applications},
        (John Wiley \& Sons, New York, 1991).
\bibitem[25]{Hussey}
        N. E. Hussey, M. N. McBrien, L. Balicas, J. S. Brooks, S. Horii,
        and H. Ikuta, Phys. Rev. Lett. {\bf 89}, 086601 (2002).
\bibitem[26]{Menzel}
        A. Menzel, R. Beer, and E. Bertel, Phys. Rev. Lett.
        {\bf 89}, 076803 (2002).
\bibitem[27]{Fuji02}
        Shin-ichi Fujimori, Akihiro Ino, Testuo Okane, Atsushi Fujimori,
        Kozo Okada, Toshio Manabe, Masahiro Yamashita, Hideo Kishida, and
        Hiroshi Okamoto, Phys. Rev. Lett. {\bf 88}, 247601 (2002).
\bibitem[28]{Hasan}
        M. Z. Hasan, P. A. Montano, E. D. Isaacs, Z.-X. Shen, H. Eisaki,
        S. K. Sinha, Z. Islam, N. Motoyama, and S. Uchida, Phys. Rev. Lett.
        {\bf 88}, 177403-1 (2002).
\bibitem[29]{Ralph}
        R. Claessen, M. Sing, U. Schwingenschl\"ogl, P. Blaha,
        M. Dressel, and C. S. Jacobsen, Phys. Rev. Lett. {\bf 88}, 096402
        (2002).
\bibitem[30]{Gweon}
        G.-H. Gweon, J. D. Denlinger, C. G. Olson, H. Hochst, J. Marcus, and
        C. Schlenker C, Physica B {\bf 312-313}, 584 (2002); G.-H. Gweon,
        J. D. Denlinger, J. W. Allen, R. Claessen, C. G. Olson, H. Hochst, J. Marcus,
        C. Schlenker, and L. F. Schneemeyer, J. Electron Spectrosc. Relat. Phenom.
        {\bf 117-118}, 481 (2001).
\bibitem[31]{Monceau}
        P. Monceau, F. Ya. Nad, and S. Brazovskii, Phys. Rev. Lett.
        {\bf 86}, 4080 (2001).
\bibitem[32]{Takenaka}
        K. Takenaka, K. Nakada, A. Osuka, S. Horii, H. Ikuta, I. Hirabayashi,
        S. Sugai, and U. Mizutani, Phys. Rev. Lett. {\bf 85}, 5428 (2000).
\bibitem[33]{Mizokawa}
        T. Mizokawa, C. Kim, Z.-X. Shen, A. Ino, T. Yoshida, A. Fujimori,
        M. Goto, H. Eisaki, S. Uchida, M. Tagami, K. Yoshida, A. I. Rykov,
        Y. Siohara, K. Tomimoto, S. Tajima, Yuh Yamada, S. Horii, N. Yamada,
        Yasuji Yamada, and I. Hirabayashi, Phys. Rev. Lett. {\bf 85}, 4779 (2000).
\bibitem[34]{Moser}
        J. Moser, J. R. Cooper, D. J\'erome, B. Alavi, S. E. Brown, and
        K. Bechgaard, Phys. Rev. Lett. {\bf 84}, 2674 (2000).
\bibitem[35]{Mihaly}
        G. Mih\'aly, I. K\'ezsm\'arki, F. Z\'amborszky, and L. Forr\'o, Phys.
        Rev. Lett. {\bf 84}, 2670 (2000).
\bibitem[36]{Vescoli00}
        V. Vescoli, F. Zwick, J. Voit, H. Berger, M. Zacchigna, L. Degiorgi,
        M. Grioni, and G. Gr\"uner, Phys. Rev. Lett. {\bf 84}, 1272
        (2000).
\bibitem[37]{Denlinger}
        J. D. Denlinger, G.H. Gweon, J. W. Allen, C. G. Olson, J. Marcus,
        C. Schlenker, and L.-S. Hsu, Phys. Rev. Lett. {\bf 82},
        2540 (1999).
\bibitem[38]{Fuji}
        H. Fujisawa, T. Yokoya, T. Takahashi, S. Miyasaka, M.
        Kibune, and H. Takagi, Phys. Rev. B {\bf 59}, 7358 (1999).
\bibitem[39]{Kobayashi}
        K. Kobayashi, T. Mizokawa, A. Fujimori, M. Isobe, Y. Ueda,
        T. Tohyama, and S. Maekawa, Phys. Rev. Lett. {\bf 82},
        803 (1999); K. Kobayashi, T. Mizokawa, and A. Fujimori, {\it ibid.}
        {\bf 80}, 3121 (1998).
\bibitem[40]{Bourbonnais}
        Claude Bourbonnais and Denis J\'erome, Science {\bf 281}, 1155 (1998).
\bibitem[41]{Vescoli}
        V. Vescoli, L. Degiorgi, W. Henderson, G. Gr\"uner, K. P. Starkey,
        and L. K. Montgomery, Science {\bf 281}, 1181 (1998).
\bibitem[42]{Zwick}
        F. Zwick, D. J\'erome, G. Margaritondo, M. Onellion, J. Voit, and
        M. Grioni, Phys. Rev. Lett. {\bf 81}, 2974 (1998);
        F. Zwick, S. Brown, G. Margaritondo, C. Merlic, M. Onellion, J. Voit,
        and M. Grioni, {\it ibid.} {\bf 79}, 3982 (1997).
\bibitem[43]{Neudert}
        R. Neudert, M. Knupfer, M. S. Golden, J. Fink, W. Stephan,
        K. Penc, N. Motoyama, H. Eisaki, and S. Uchida, Phys. Rev. Lett.
        {\bf 81}, 657 (1998).
\bibitem[44]{Mori}
        T. Mori, T. Kawamoto, J. Yamaura, T. Enoki, Y. Misaki, T. Yamabe,
        H. Mori, and S. Tanaka, Phys. Rev. Lett. {\bf 79}, 1702 (1997).
\bibitem[45]{Kim}
        C. Kim, A. Y. Matsuura, Z.-X. Shen, N. Motoyama, H. Eisaki, S. Uchida,
        T. Tohyama, and S. Maekawa, Phys. Rev. Lett. {\bf 77}, 4054
        (1996); N. Motoyama, H. Eisaki, and S. Uchida, Phys. Rev. Lett. {\bf 76},
        3212 (1996).
\bibitem[46]{Luther}
        A. Luther and I. Peschel, Phys. Rev. B {\bf 9}, 2911
        (1974); {\bf 12}, 3908 (1975).
\bibitem[47]{Solyom}
        J. S\'olyom, Adv. Phys. {\bf 28}, 201 (1979).
\bibitem[48]{Haldane}
        F. D. M. Haldane, J. Phys. C {\bf 14},
        2585 (1981).
\bibitem[49]{Faddeed}
        L. D. Faddeev and L. A. Takhtajan, Phys. Lett. {\bf 85A},
        375 (1981).
\bibitem[50]{PWA}
        J. C. Talstra, S. P. Strong, and P. W. Anderson, Phys. Rev. Lett. {\bf 74},
        5256 (1995).
\bibitem[51]{Talstra}
        J. C. Talstra and S. P. Strong, Phys. Rev. B {\bf 56}, 6094 (1997).
\bibitem[52]{Tohyama}
        Y. Mizuno, K. Tsutsui, T. Tohyama, and S. Maekawa, Phys. Rev.
        B {\bf 62}, R4769 (2000);
        H. Kishida, M. Ono, K. Miura, H. Okamoto, M. Izumi, T. Manako,
        M. Kawasaki, Y. Taguchi, Y. Tokura, T. Tohyama, K. Tsutsui, and
        S. Maekawa, Phys. Rev. Lett. {\bf 87}, 177401 (2001).
\bibitem[53]{Granath}
        M. Granath, V. Oganesyan, D. Orgad, and S. A. Kivelson,
        Phys. Rev. B {\bf 65}, 184501 (2002).
\bibitem[54]{Orgard}
        D. Orgad, S. A. Kivelson, E. W. Carlson, V. J. Emery,
        X. J. Zhou, and Z. X. Shen, Phys. Rev. Lett. {\bf 86},
        4362 (2001).
\bibitem[55]{Carlson}
        E. W. Carlson, D. Orgad, S. A. Kivelson, and  V. J. Emery,
        Phys. Rev. B {\bf 62}, 3422 (2000).
\bibitem[56]{Zaanen}
        J. Zaanen, O. Y. Osman, H. V. Kruis, Z. Nussinov, and J.
        Tworzydlo, Philos. Magaz. B {\bf 81}, 1485 (2001).
\bibitem[57]{Antonio}
        A. L. Chernyshev, S. R. White, and A. H. Castro Neto, Phys.
        Rev. B {\bf 65}, 214527 (2002).
\bibitem[58]{spectral0}
        M. Sing, U. Schwingenschl\"ogl, R. Claessen, P.
        Blaha, J. M. P. Carmelo, L. M. Martelo, P. D. Sacramento, M.
        Dressel, and C. S. Jacobsen, submitted for publication in Phys. Rev. B (2003)
        [cond-mat/0304283].
\bibitem[59]{Essler}
        Fabian H. L. Essler, Vladimir E. Korepin, and
        Kareljan Schoutens, Phys. Rev. Lett. {\bf 67}, 3848 (1991);
        Nucl. Phys. B {\bf 372}, 559 (1992).
\bibitem[60]{IIIb}
        J. M. P. Carmelo, submitted for publication
        in Phys. Rev. B (2003) [cond-mat/0305568].
\bibitem[61]{V}
        J. M. P. Carmelo and K. Penc, to be submitted for publication
        in Phys. Rev. B (2003).
\bibitem[62]{spectral}
        J. M. P. Carmelo, K. Penc, L. M. Martelo, P. D. Sacramento,
        J. M. B. Lopes dos Santos, R. Claessen, M. Sing, and
        U. Schwingenschl\"ogl, to be submitted for publication
        (2003).
\bibitem[63]{optical}
        J. M. P. Carmelo, N. M. R. Peres, and P. D.
        Sacramento, Phys. Rev. Lett. {\bf 84}, 4673 (2000).
\bibitem[64]{dynamical}
        J. M. P. Carmelo, K. Penc, and P. D. Sacramento,
        to be submitted for publication (2003).
\bibitem[65]{phases}
        J. M. P. Carmelo, F. Guinea, and P. D. Sacramento,
        to be submitted for publication (2003).
\bibitem[66]{IV}
        J. M. P. Carmelo, L. M. Martelo, and P. D. Sacramento, to be
        submitted for publication (2003).
\bibitem[67]{Gu}
        S. J. Gu, N. M. R. Peres, and J. M. P. Carmelo, to be submitted
        for publication (2003).
\bibitem[68]{Schulz}
        H. Schulz, Phys. Rev. Lett. {\bf 64}, 2831 (1990).
\bibitem[69]{takahashi77}
        M. Takahashi, J. Phys. C {\bf 10}, 1289 (1977).
\bibitem[70]{Belavin}
        A. A. Belavin, A. M. Polyakov, and A. B.
        Zamolodchikov, J. Stat. Phys. {\bf 34},
        763 (1984); Nucl. Phys. B {\bf 241}, 333 (1984);
        Holger Frahm and V. E. Korepin, Phys. Rev. B {\bf 42},
        10 553 (1990); {\it ibid.} {\bf 43}, 5653 (1991).
\bibitem[71]{Senechal}
        D. S\'en\'echal, D. Perez, and M. Pioro-Ladri\`ere,
        Phys. Rev. Lett. {\bf 84}, 522 (2000).
\bibitem[72]{Controzzi}
        D. Controzzi, F. H. L. Essler, and A. M. Tsvelik,
        Phys. Rev. Lett. {\bf 86}, 680 (2001).
\end{references}
\end{document}